\documentclass[aps,physrev,reprint,groupedaddress]{revtex4-2}
\usepackage{graphicx,bm,amsmath,amssymb,natbib,url,epsfig}
\usepackage{dcolumn}
\usepackage[colorlinks=true,linkcolor=blue,allcolors=blue]{hyperref}
\usepackage{enumerate}
\usepackage{array}
\newcolumntype{C}[1]{>{\centering\arraybackslash}p{#1}}

\usepackage{color}
\usepackage{xcolor}
\usepackage[normalem]{ulem}
\definecolor{darkgreen}{rgb}{0.1,.6,.1}

\definecolor{tan1}{HTML}{ffa040}
\definecolor{brown4}{HTML}{801414}
\definecolor{light_salmon}{HTML}{ffa070}

\newcommand{\ohm}{\ensuremath{\Omega}}

\begin{document}


\title{Emergence of minimal chimera in uncoupled oscillators under common frequency-modulated driving: Theory and experiment}


\author{Debabrata Biswas${}^{1}$}
\thanks{Corresponding author}
\email{debbisrs@gmail.com}
\author{Tanmoy Banerjee${}^{2}$}

\affiliation{\vspace{0.1in} ${}^{1}$Department of Physics, Bankura University, Bankura 722 155, West Bengal, India\\  ${}^{2}$Chaos and Complex Systems Research
  Laboratory, Department of Physics, University of Burdwan, Burdwan
  713 104, West Bengal, India\\ }%
\received{:to be included by reviewer}
\date{\today}

\begin{abstract}
We report the experimental realization of minimal chimera states in a system of three uncoupled oscillators driven solely by frequency-modulated forcing. Unlike conventional scenarios where chimera states emerge due to interactions among oscillators, here the coexistence of coherent and incoherent dynamics arises entirely from a common external modulation of a system parameter. By tuning the modulation amplitude and frequency, the system exhibits transitions between global synchronization, global incoherence, and minimal chimera states. The stability of these regimes is quantified using the maximal Lyapunov exponent, while a synchronization order parameter is employed to characterize the degree of coherence. A systematic exploration of the parameter space reveals well-defined regions associated with distinct dynamical behaviors. To provide analytical understanding, we employ a phase-reduction approach and derive the corresponding phase dynamics, which elucidate the mechanisms underlying phase locking and desynchronization. The robustness of the proposed mechanism is further demonstrated in a time-delayed chaotic system. Finally, experimental results obtained from an electronic circuit realization confirm the emergence of minimal chimera states under frequency-modulated driving. These findings establish external modulation as a viable route to chimera formation without coupling, offering a new perspective on collective dynamics in driven nonlinear systems.

\end{abstract}


\maketitle

\section{Introduction}
\label{sec:intro}

The chimera, a mythical hybrid creature, has long served as a metaphor for the coexistence of contrasting features. In nonlinear dynamics, chimera states describe spatiotemporal patterns in networks of identical oscillators where coherent (synchronized) and incoherent (desynchronized) subpopulations coexist. Since their first observation in nonlocally coupled phase oscillators \cite{kuramoto2002coexistence, abrams2004chimera, panaggio2015chimera}, chimera states have attracted sustained interest due to their counterintuitive nature and broad applicability. Early studies suggested that such states require large network sizes and specially prepared initial conditions \cite{panaggio2015chimera}, which delayed their experimental realization. This gap was later bridged through demonstrations in systems such as spatial light modulators \cite{hagerstrom2012experimental} and chemical oscillators \cite{tinsley2012chimera}.

Subsequent investigations have revealed chimera states across a wide range of physical and engineered systems, including electrochemical oscillators \cite{schmidt2014coexistence}, Lorenz systems \cite{meena2015chimera}, coupled metronomes \cite{martens2013chimera}, optical frequency combs \cite{viktorov2014coherence}, and various electronic and optoelectronic platforms \cite{larger2013virtual, rosin2014transient, larger2015laser, hart2016experimental}. Moreover, it is now well established that chimeras can arise in relatively small networks \cite{meena2015chimera, ashwin2015weak, hart2016experimental}, under random initial conditions \cite{sethia2014chimera, nkomo2013chimera}, and even in globally coupled configurations \cite{sethia2014chimera, yeldesbay2014chimeralike, schmidt2015clustering, schmidt2015chimeras}. These findings have significantly expanded the conceptual framework of chimera formation.

In parallel, synchronization induced by common external inputs has emerged as an alternative route to collective behavior. Traditionally, synchronization was attributed to direct coupling between oscillators \cite{pikovsky2001synchronization}. However, it is now well known that shared external forcing or noise can synchronize uncoupled systems \cite{gammaitoni1995stochastic, gonzalez2008forced}. This phenomenon has been extensively studied in diverse contexts, including limit-cycle oscillators subjected to random impulses \cite{nakao2005synchrony, Nakao2007}, biological oscillators influenced by external signals \cite{chen2010multicellular}, and stochastic or noisy environments such as laser systems \cite{wieczorek2011noise}, bistable systems \cite{rozenfeld2001noise}, and chemical oscillators \cite{miyakawa2002noise}. Noise-induced synchronization has also been reported in chaotic systems \cite{zhou2002noise, Guan2006, lai2010route} and neural networks \cite{Lang2010, Shuai1998}, with experimental confirmations \cite{Barbay2003, Vaidya2021} and extensions to colored noise scenarios \cite{wang2009onset}. More recently, frequency-modulated driving has been shown to significantly influence the stability and synchronization properties of dynamical systems, including phase oscillators, Van der Pol oscillators, and ecological models \cite{lucas2018stabilization, lucas2019nonautonomous, bhandary23stability}. Noise can also play a nontrivial role in the emergence and persistence of chimera states. Previous studies have shown that stochastic perturbations may induce, sustain, or modify chimera-like dynamics in oscillator networks \cite{Semenova2016, Zhu2023, Khatun2023}. These results further indicate that partial synchronization is not necessarily restricted to ideal deterministic systems and may survive, or even arise from, fluctuations.

Alongside these developments, the study of minimal chimera states, defined as the smallest network configurations exhibiting coexistence of coherence and incoherence, has gained increasing attention. Several configurations have been identified, including systems with two synchronized and one desynchronized oscillator, as well as mixed periodic–chaotic states in coupled units \cite{hart2016experimental, wojewoda2016smallest, awal2019smallest}. Experimental realizations have been reported in optoelectronic networks \cite{hart2016experimental}, mechanical oscillators \cite{wojewoda2016smallest}, R\"{o}ssler systems \cite{meena2015chimera}, and pendulum-like networks \cite{maistrenko2017smallest}. Additional examples include metronomes with phase-lagged coupling \cite{ebrahimzadeh2020minimal}, electrochemical oscillators \cite{wiehl2021birhythmicity}, and Josephson junction arrays \cite{saha2021smallest}. Notably, all previously reported minimal chimera states rely on direct or indirect coupling among the oscillators.

This raises a fundamental question: can chimera states, particularly minimal chimeras, emerge in completely uncoupled systems solely under the influence of external driving? Addressing this issue is crucial for understanding whether the coexistence of coherent and incoherent dynamics is intrinsically linked to coupling, or whether it can arise as a purely driven phenomenon. However, this issue has not been addressed yet.

In this work, we demonstrate that minimal chimera states can indeed be induced in a system of three uncoupled oscillators through frequency-modulated driving of a system parameter. We consider systems ranging from limit-cycle dynamics to chaotic regimes, and show that a common frequency-modulation can generate heterogeneous dynamical responses among otherwise identical units, leading to the coexistence of synchronized and desynchronized behavior. By systematically varying both intrinsic system parameters and the characteristics of the modulation, we uncover a rich variety of dynamical regimes, including global synchrony, global incoherence, and multiple chimera configurations.

To characterize these states, we compute the maximal Lyapunov exponent to assess dynamical stability and employ a synchronization order parameter to quantify coherence. Furthermore, we apply a phase reduction approach to derive reduced phase equations, which provide analytical insight into phase locking and desynchronization mechanisms. To demonstrate the generality of the proposed mechanism, we extend our analysis to a time-delayed chaotic system and observe similar chimera behavior under frequency-modulated forcing. Finally, we validate our theoretical predictions through an electronic circuit realization, where minimal chimera states are experimentally observed. Lastly, in Appendix A, we apply the present scheme to a non-delayed chaotic R\"{o}ssler system and obtain similar scenarios.


\section{Model-I: Van der Pol oscillator with frequency-modulated driving}
\label{sec:mod1vdp}

We begin with the classical Van der Pol oscillator, described by
\begin{equation}\label{vdpeq}
    \ddot{x}-\mu(1-x^2)\dot{x}+\omega^2 x=0,
\end{equation}
where $\mu \in \mathbb{R}$ controls the strength of nonlinearity and $\omega$ denotes the natural frequency of the oscillator.

To incorporate external driving, we introduce a frequency-modulation in the nonlinear parameter $\mu$ \cite{lucas2018stabilization, lucas2019nonautonomous, bhandary23stability}. Specifically, we consider

\begin{equation}\label{mueq}
    \mu\rightarrow \mu\big(1+r\sin(\vartheta_0)\big),
\end{equation}
where $r>0$ represents the amplitude of the modulation and $\vartheta_0$ is its phase.

The phase $\vartheta_0$ evolves according to a frequency-modulated signal, given by
\begin{equation}\label{phieq}
    \frac{d\vartheta_0}{dt}=\omega_0 \big(1+\delta F(\omega_f t)\big),
\end{equation}
where the function $F(\cdot)$ is periodic function with single tone having the form $F(\omega_f t)=\sin(\omega_f t)$. The non-modulated (carrier) frequency is given by $\omega_0\in\mathcal{R}^+$. The strength and frequency of the perturbation are given by $\delta\in\mathcal{R}^+$ and $\omega_f\in\mathcal{R}^+$, respectively. The phase of the forcing frequency evolves as $\vartheta_0=\omega_0\big(t-\frac{\delta}{\omega_f}\cos(\omega_f t)\big)$.

Next, we consider a network of three van der Pol oscillators driven by a common frequency-modulated drive
\begin{equation}\label{vdpeqi}
    \ddot{x}_i-\mu\big(1+r\sin(\vartheta_0)\big)(1-x_i^2)\dot{x}_i+\omega^2 x_i=0.
\end{equation}
Here, $i=1,2,3$ represents the indices of the oscillators. 

\begin{figure}
    \centering
    \includegraphics[width=0.4\textwidth]{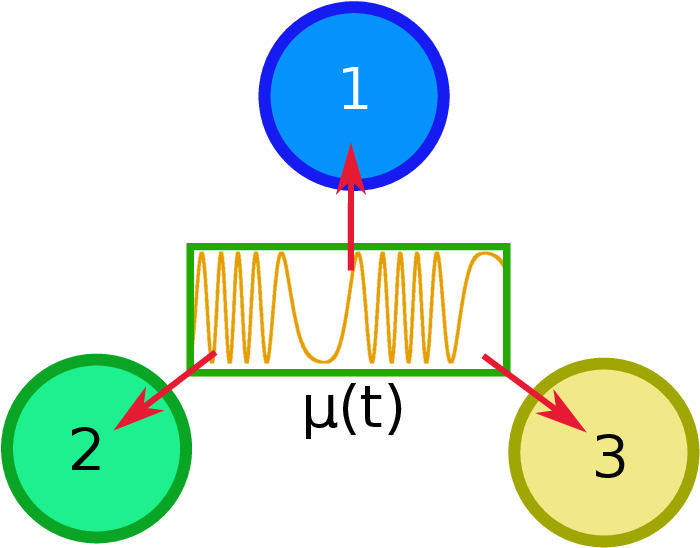}
    \caption{Schematic representation of three identical oscillators driven by a common frequency-modulated forcing.}
    \label{schema}
\end{figure}

In the present setup, nonautonomous dynamics are introduced through the modulation parameter $r \neq 0$, which acts as a common external drive for all oscillators. We consider three identical Van der Pol oscillators initialized with different initial conditions, as illustrated schematically in Fig.~\ref{schema}.

In the absence of modulation $(r=0)$, the system reduces to the standard autonomous van der Pol oscillator exhibiting a stable limit cycle. However, for $r\neq 0$, the system becomes nonautonomous due to the frequency-modulated driving, which introduces a common external forcing across all oscillators. 

\begin{figure}
    \centering
    \includegraphics[width=0.5\textwidth]{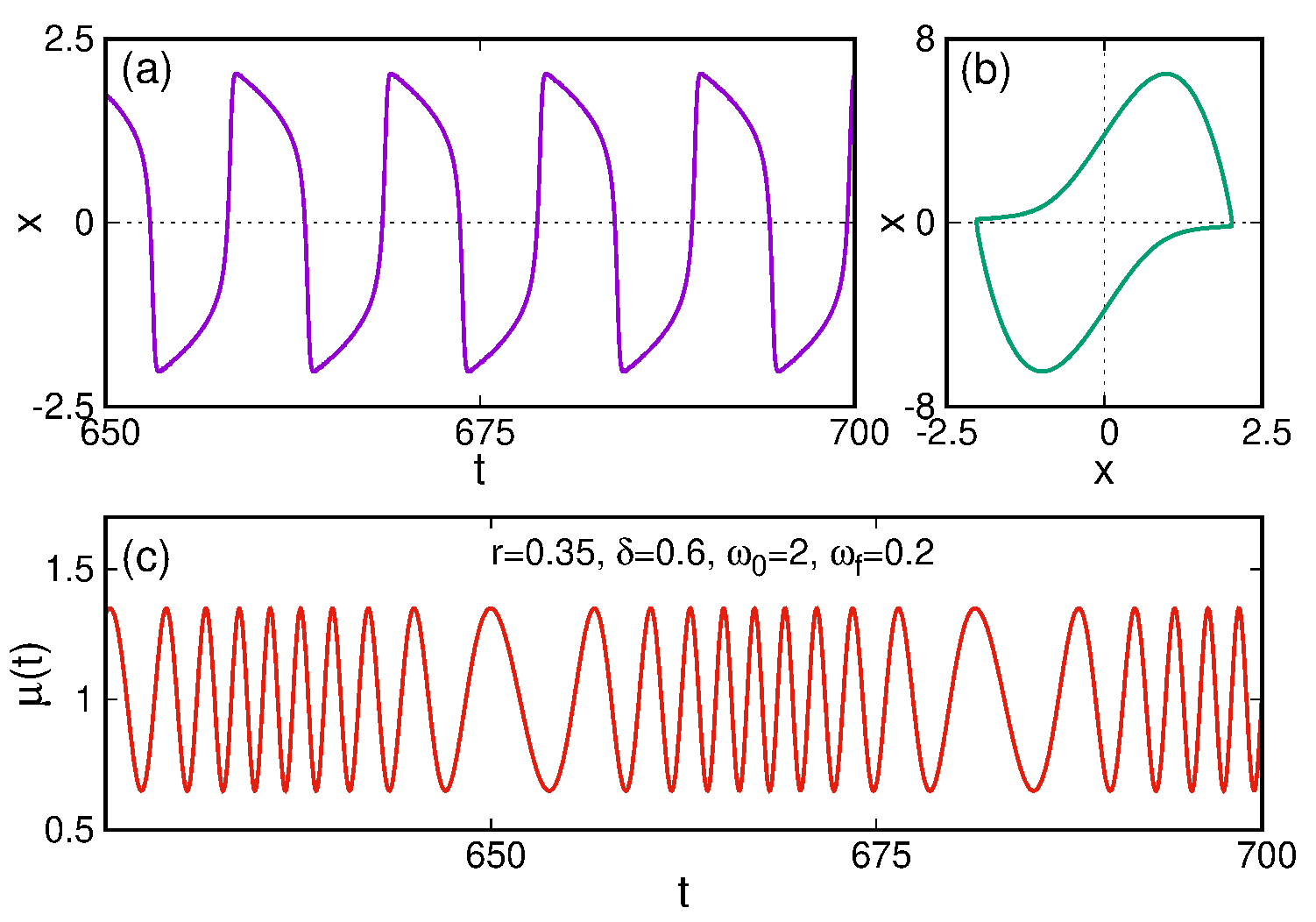}
    \caption{(a) Time series and (b) phase portrait of the autonomous (r = 0) Van der Pol oscillator for $\mu = 4.1$. (c) Time evolution of the frequency-modulated parameter $\mu(t)$ for $r = 0.35$, $\delta = 0.6$, $\omega_0 = 2$, and $\omega_f = 0.2$.}
    \label{vdp_ts}
\end{figure}

The time series and phase plane plots of the autonomous ($\mu(t)\rightarrow \mu$) Van der Pol oscillator are shown in Fig.~\ref{vdp_ts}(a) and (b), respectively.
The frequency-modulated waveform of $\mu(t)$ is shown in Fig.~\ref{vdp_ts}(c) for an exemplary set of parameters: $r=0.35$, $\delta=0.6$, $\omega_0=2$, and $\omega_f=0.2$.

We show below that this external modulation plays a crucial role in generating a variety of collective dynamical behaviors, including synchronization and minimal chimera states.

\begin{figure}
    \centering
    \includegraphics[width=0.48\textwidth]{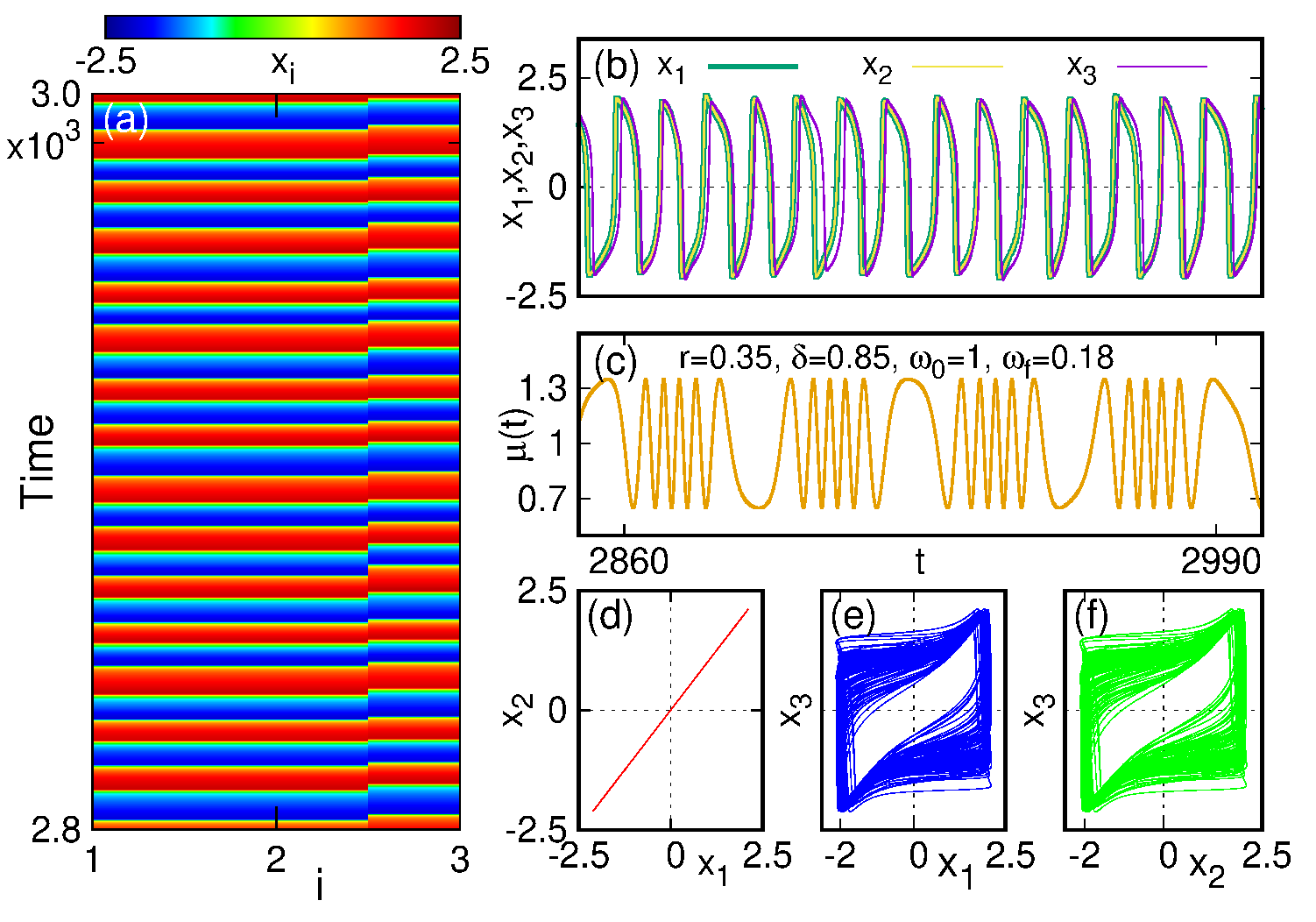}
    \caption{Minimal chimera state in the frequency-modulated Van der Pol system. (a) Spatiotemporal evolution of the oscillators. (b) Corresponding time series. (c) Frequency-modulated driving $\mu(t)$. (d) Phase portrait in the $x_1$–$x_2$ plane showing complete synchronization. (e),(f) Phase portraits in the $x_1$–$x_3$ and $x_2$–$x_3$ planes showing desynchronization. Parameters: $\mu = 4.1$, $r = 0.35$, $\delta = 0.85$, $\omega_0 = 1.0$, $\omega_f = 0.18$, and $\omega = 1.0$.}
    \label{vdp_fm_sptmp}
\end{figure}

\subsection{Results: Minimal chimera in a network of frequency modulated parameter driven Van der Pol oscillators}
\label{sub:resvdp}


In the absence of modulation ($r = 0$), the system remains autonomous, and the oscillators evolve independently, showing no tendency toward synchronization. However, once the modulation is activated ($r \neq 0$), the common frequency-modulated forcing induces nontrivial collective behavior. Depending on the parameter values, the system exhibits several synchronization patterns: complete synchronization, complete incoherence, or partial synchronization in which a subset of oscillators synchronize while others remain desynchronized. The latter corresponds to a minimal chimera state. In particular, we observe parameter regimes where two oscillators evolve in synchrony, while the third remains incoherent.

Figure~\ref{vdp_fm_sptmp} illustrates a representative example of the minimal chimera pattern for $r = 0.35$, $\delta = 0.85$, $\omega_0 = 1.0$, $\omega_f = 0.18$, and $\mu = 4.1$, the spatiotemporal evolution [Fig.~\ref{vdp_fm_sptmp}(a)] reveals that $x_1$ and $x_2$ follow identical trajectories, whereas $x_3$ evolves independently. This observation is corroborated by the time series in Fig.~\ref{vdp_fm_sptmp}(b), where $x_1$ and $x_2$ overlap, while $x_3$ deviates significantly. The corresponding modulation parameter $\mu(t)$ is shown in Fig.~\ref{vdp_fm_sptmp}(c). Further insight is obtained from the phase-space projections. The $x_1$–$x_2$ projection [Fig.~\ref{vdp_fm_sptmp}(d)] collapses onto a diagonal line, indicating complete synchronization between these oscillators. In contrast, the projections $x_1$–$x_3$ and $x_2$–$x_3$ [Fig.~\ref{vdp_fm_sptmp}(e,f)] exhibit scattered trajectories, confirming the absence of synchronization involving $x_3$. Together, these results establish the occurrence of a minimal chimera state in the system.

\begin{figure*}
    \centering
    \includegraphics[width=0.8\textwidth]{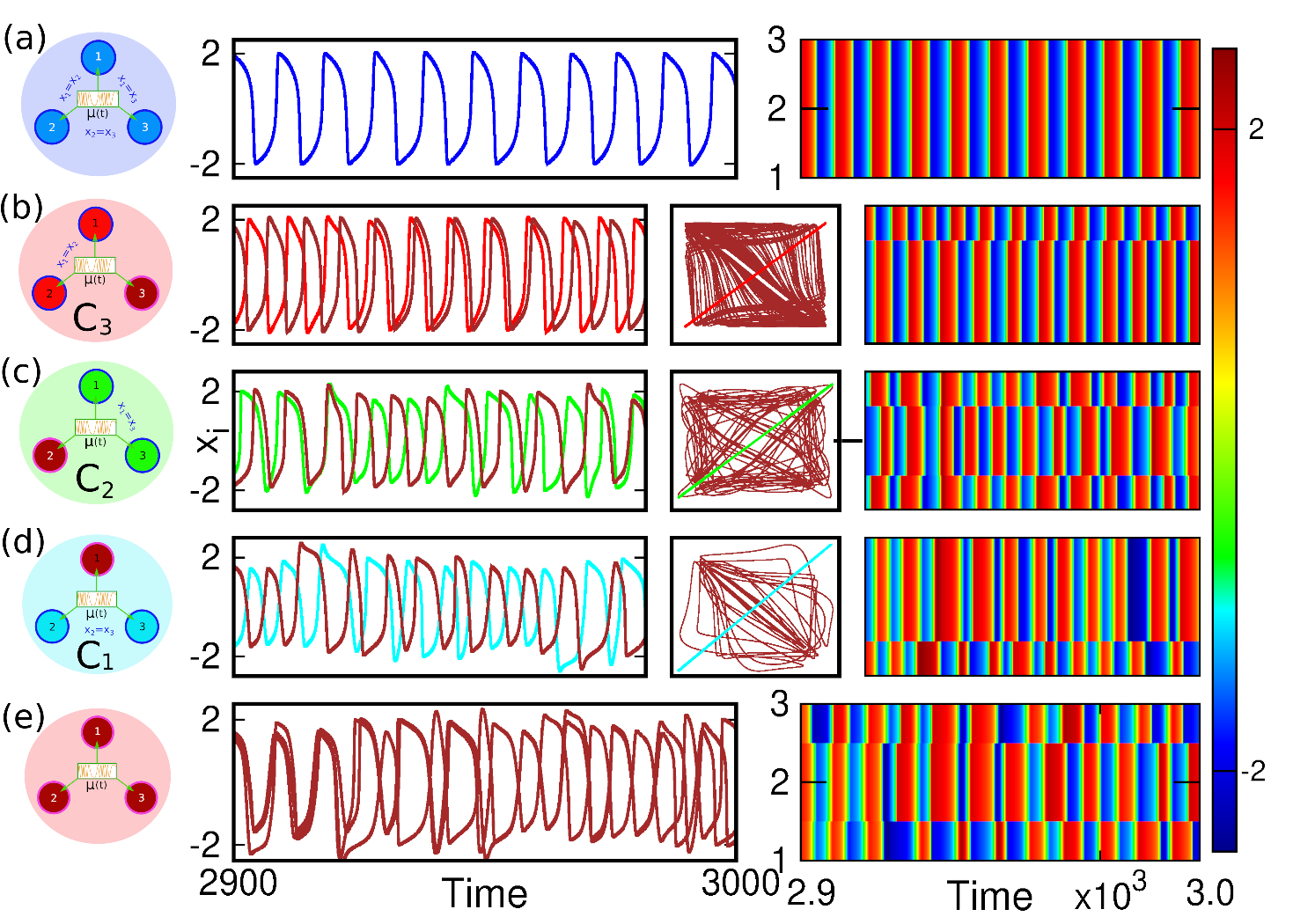}
    \caption{Representative dynamical states of the Van der Pol system. Left column: schematic representation of synchronization patterns. Middle column: time series. Right column: spatiotemporal evolution. Insets: phase-plane plots. (a) Global synchronization ($\mu = 3.75$, $r = 0.2$). (b) Chimera state C$_3$ ($x_1$–$x_2$ synchronized, $x_3$ desynchronized). (c) Chimera state C$_2$. (d) Chimera state C$_1$. (e) Global incoherence ($\mu = 4.06$, $r = 0.94$). Other parameters: $\delta = 0.5$, $\omega_0 = 1.0$, $\omega_f = 0.1$, $\omega = 1.0$.}
    \label{vdp_sync_usync}
\end{figure*}

To systematically characterize the different dynamical regimes, we examine the temporal evolution of the oscillators for a range of parameters, as shown in Fig.~\ref{vdp_sync_usync}. As initial conditions we use: $(x_1(0), y_1(0))=(0.9,0.4)$, $(x_2(0), y_2(0))=(0.7,0.3)$, and $(x_3(0), y_3(0))=(0.5,0.65)$ throughout the investigation. Here, we fix $\delta = 0.5$, $\omega_0 = 1.0$, $\omega_f = 0.1$, and $\omega = 1.0$, and vary $\mu$ and $r$. The left column presents schematic representations of the states, the middle column shows the time series, and the right column displays the corresponding spatiotemporal patterns. The insets show the phase-plane plots.

For $\mu = 3.75$ and $r = 0.2$, the system exhibits global synchronization [Fig.~\ref{vdp_sync_usync}(a)], where all oscillators evolve identically. In contrast, Fig.~\ref{vdp_sync_usync}(b) shows a minimal chimera configuration (denoted as C$_3$), in which $x_1$ and $x_2$ are synchronized while $x_3$ remains incoherent (see Subsec.~\ref{subsec:vdpmur} and Table-\ref{tab1} for the definitions of the collective states).

A complementary configuration, labeled C$_2$, is shown in Fig.~\ref{vdp_sync_usync}(c), where $x_1$ and $x_3$ synchronize and $x_2$ behaves incoherently. Similarly, Fig.~\ref{vdp_sync_usync}(d) presents the C$_1$ state, characterized by synchronization between $x_2$ and $x_3$, with $x_1$ remaining desynchronized. Finally, for $\mu = 4.06$ and $r = 0.94$, the system exhibits complete incoherence [Fig.~\ref{vdp_sync_usync}(e)], where no pair of oscillators displays synchronized behavior.

These observations demonstrate that the frequency-modulated driving induces a wide variety of collective states in the absence of direct coupling. The specific dynamical regime realized depends sensitively on the interplay between the intrinsic system parameters and the modulation characteristics.

\begin{figure*}
    \centering
    \includegraphics[width=0.7\textwidth]{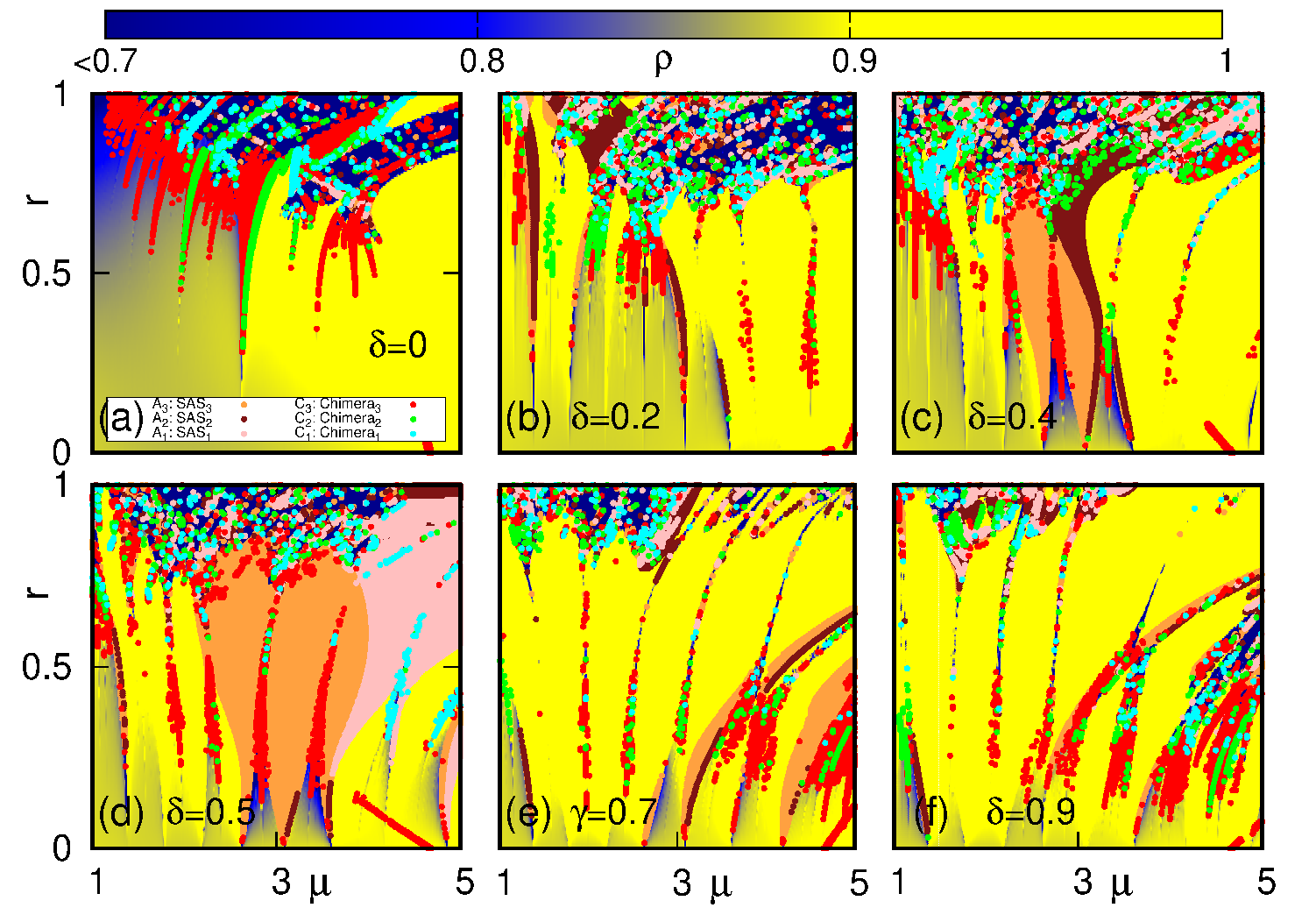}
    \caption{Synchronization landscapes of the Van der Pol system in the $\mu$–$r$ plane for different values of $\delta$: (a) $\delta = 0$, (b) $\delta = 0.2$, (c) $\delta = 0.4$, (d) $\delta = 0.5$, (e) $\delta = 0.7$, and (f) $\delta = 0.9$. Colors indicate the order parameter $\rho$, while markers denote chimera (C$_i$) and CS–Anti-S (A$_i$) states (see Table-\ref{tab1}).}
    \label{vdp_two_par}
\end{figure*}

The emergence of chimera states in the present system can be understood in terms of an effective interaction induced by the common external modulation. Although the oscillators are not directly coupled, the shared frequency-modulated forcing acts as a global drive that influences all units simultaneously. Due to differences in initial conditions and nonlinear responses, the oscillators do not respond identically to this driving. As a result, subsets of oscillators may lock to the modulation in a coherent manner, while others fail to synchronize, leading to the coexistence of coherent and incoherent groups. In this sense, the common forcing plays a role analogous to effective coupling, giving rise to clustering and minimal chimera states even in the absence of explicit interactions.

\subsection{Dynamics of of the network in $\mu-r$ space}
\label{subsec:vdpmur}

In this subsection, we investigate the collective dynamics of the system in the $\mu$--$r$ parameter space for different values of the modulation depth $\delta$. To quantify the degree of synchronization, we employ the synchronization order parameter $\rho$ \cite{Komin2011Synchronization}, which provides a global measure of coherence among the oscillators. This quantity is defined as \cite{bhandary23stability, Komin2011Synchronization}
\begin{equation}\label{rho}
    \rho=\sqrt{\left\langle\frac{\bar{z}(t)^2}{\frac{1}{N}\sum_{i=1}^{N}z_i(t)^2}\right\rangle},
\end{equation}
where $z_i(t)$ denotes the state variable of the $i$-th oscillator at time $t$, $\bar{z}(t) = \frac{1}{N}\sum_{i=1}^{N} z_i(t)$ is the ensemble average, and $\langle \cdot \rangle$ represents time averaging over a sufficiently long interval. The order parameter satisfies $0 \leq \rho \leq 1$, with $\rho = 1$ corresponding to complete synchronization, $\rho = 0$ indicating complete incoherence, and intermediate values representing partial synchronization.

Using this measure, we compute the synchronization landscape in the $\mu$--$r$ plane for different values of $\delta$, as shown in Fig.~\ref{vdp_two_par}. Throughout this analysis, we fix $\omega_0 = 1.0$, $\omega_f = 0.1$, and $\omega = 1.0$. The color scale represents the magnitude of $\rho$, while different dynamical states are marked by colored symbols. For $\delta = 0$ [Fig.~\ref{vdp_two_par}(a)], several distinct regimes can be identified. The red (\textcolor{red}{\Large\textbullet}), green (\textcolor{green}{\Large\textbullet}), and cyan (\textcolor{cyan}{\Large\textbullet}) markers correspond to the chimera states C$_3$, C$_2$, and C$_1$, respectively, indicating configurations where two oscillators are synchronized, and the third remains incoherent. Particularly, C$_3$ is the chimera state where $x_1$ and $x_2$ are coherent and $x_3$ is incoherent, and so on. The definitions of C$_i$ are summarized in Table-\ref{tab1}. Since the oscillators are identical and the system is permutation symmetric, the configurations $C_1$, $C_2$, and $C_3$ are symmetry-related realizations of the same minimal chimera state. They differ only in the identity of the incoherent oscillator \cite{hart2016experimental}. Similarly, $A_1$, $A_2$, and $A_3$ denote symmetry-related configurations of the same collective state. The subscripts are retained only to identify the oscillator configuration and to facilitate the presentation of the numerical results.

In addition to these chimera states, we observe regimes characterized by the coexistence of complete synchronization (CS) and anti-synchronization (Anti-S). These states are labeled as A$_3$ [denoted by the (\textcolor{pink}{\Large\textbullet}) pink dots, A$_2$ [(\textcolor{brown4}{\Large\textbullet}) brown dots], and A$_1$ [(\textcolor{tan1}{\Large\textbullet}) tan dots], where one oscillator evolves in anti-phase relative to a synchronized pair (Table-\ref{tab1}). Specifically, A$_3$ corresponds to synchronization between $x_1$ and $x_2$ with $x_3$ in anti-synchrony, and so on. These states are indicated by distinct colored markers in the figure. The corresponding time series are not shown here for brevity.

\begin{table*}
\caption{The different collective states observed in the system are classified as follows:}
\label{tab1}
\centering 
\begin{tabular}{|C{2cm}|C{2cm}|C{4cm}|C{5cm}|}
\hline
\textbf{State} & \textbf{Symbol} & \textbf{Synchronized pair} & \textbf{Remaining oscillator} \\
\hline
C$_1$ & \textcolor{cyan}{\Large\textbullet} & $(x_2, x_3)$ & $x_1$ (desynchronized) \\
C$_2$ & \textcolor{green}{\Large\textbullet} & $(x_1, x_3)$ & $x_2$ (desynchronized) \\
C$_3$ & \textcolor{red}{\Large\textbullet} & $(x_1, x_2)$ & $x_3$ (desynchronized) \\
\hline
A$_1$ & \textcolor{tan1}{\Large\textbullet} & $(x_2, x_3)$ & $x_1$ (anti-synchronized) \\
A$_2$ & \textcolor{brown4}{\Large\textbullet} & $(x_1, x_3)$ & $x_2$ (anti-synchronized) \\
A$_3$ & \textcolor{pink}{\Large\textbullet} & $(x_1, x_2)$ & $x_3$ (anti-synchronized) \\
\hline
\end{tabular}
\end{table*}

Figures~\ref{vdp_two_par}(b--f) illustrate the evolution of these dynamical regimes as $\delta$ increases from $0.2$ to $0.9$. A progressive reorganization of the parameter space is observed, with notable changes in the distribution of synchronized, chimera, and incoherent states.

\begin{figure}
    \centering
    \includegraphics[width=0.5\textwidth]{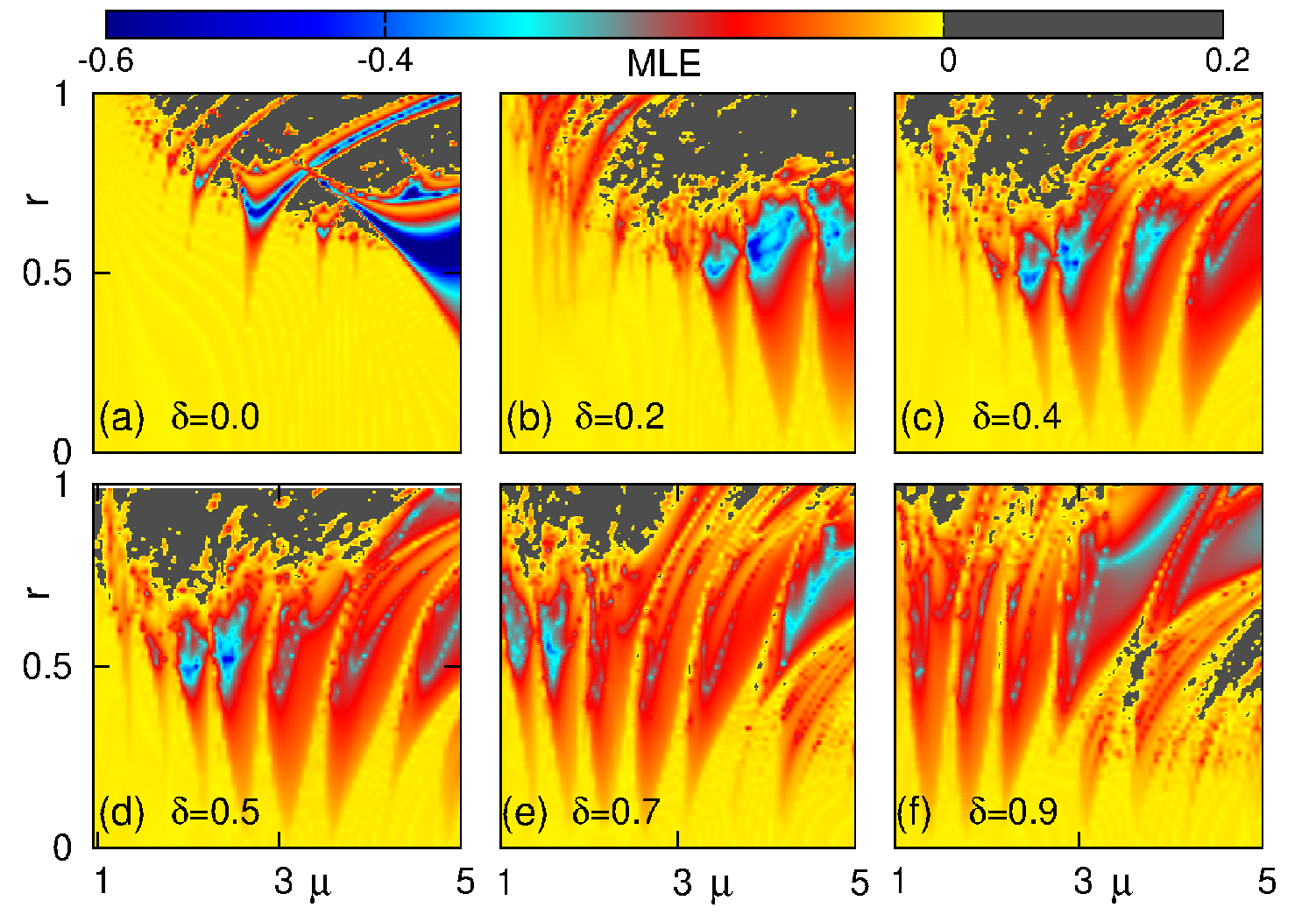}
    \caption{Maximal Lyapunov exponent (MLE) in the $\mu$–$r$ parameter space of Van der Pol system for (a) $\delta = 0$, (b) $\delta = 0.2$, (c) $\delta = 0.4$, (d) $\delta = 0.5$, (e) $\delta = 0.7$, and (f) $\delta = 0.9$. Dark gray regions correspond to incoherent dynamics. Other parameters are the same as in Fig.~\ref{vdp_two_par}.}
    \label{vdp_mle}
\end{figure}

To further assess the stability of these regimes, we compute the maximal Lyapunov exponent (MLE) across the same parameter space, as shown in Fig.~\ref{vdp_mle}. The panels correspond to the same set of $\delta$ values as in Fig.~\ref{vdp_two_par}. Regions with negative MLE indicate stable dynamics, while positive values correspond to chaotic or unstable behavior. The dark gray regions highlight parameter regimes associated with incoherent dynamics. Notably, as $\delta$ increases, the extent of these incoherent regions diminishes, indicating that stronger modulation tends to promote synchronization in the system. 

This trend suggests that frequency modulation acts as an effective organizing mechanism, enhancing coherence among otherwise uncoupled oscillators.

In Appendix-\ref{app:ic}, we dicsuss the effect of initial conditions on the system state space.

\subsection{Effect of variation of Frequency Modulation parameters}
\label{subsec:vdpotherpars}

In this subsection, we examine how the collective dynamics depend on the parameters governing the frequency-modulated driving, namely $r$, $\delta$, $\omega_0$, and $\omega_f$. For this analysis, the intrinsic parameters are fixed at $\mu = 3.0$ and $\omega = 1.0$, while the modulation parameters are varied systematically.

\begin{figure}
    \centering
    \includegraphics[width=0.5\textwidth]{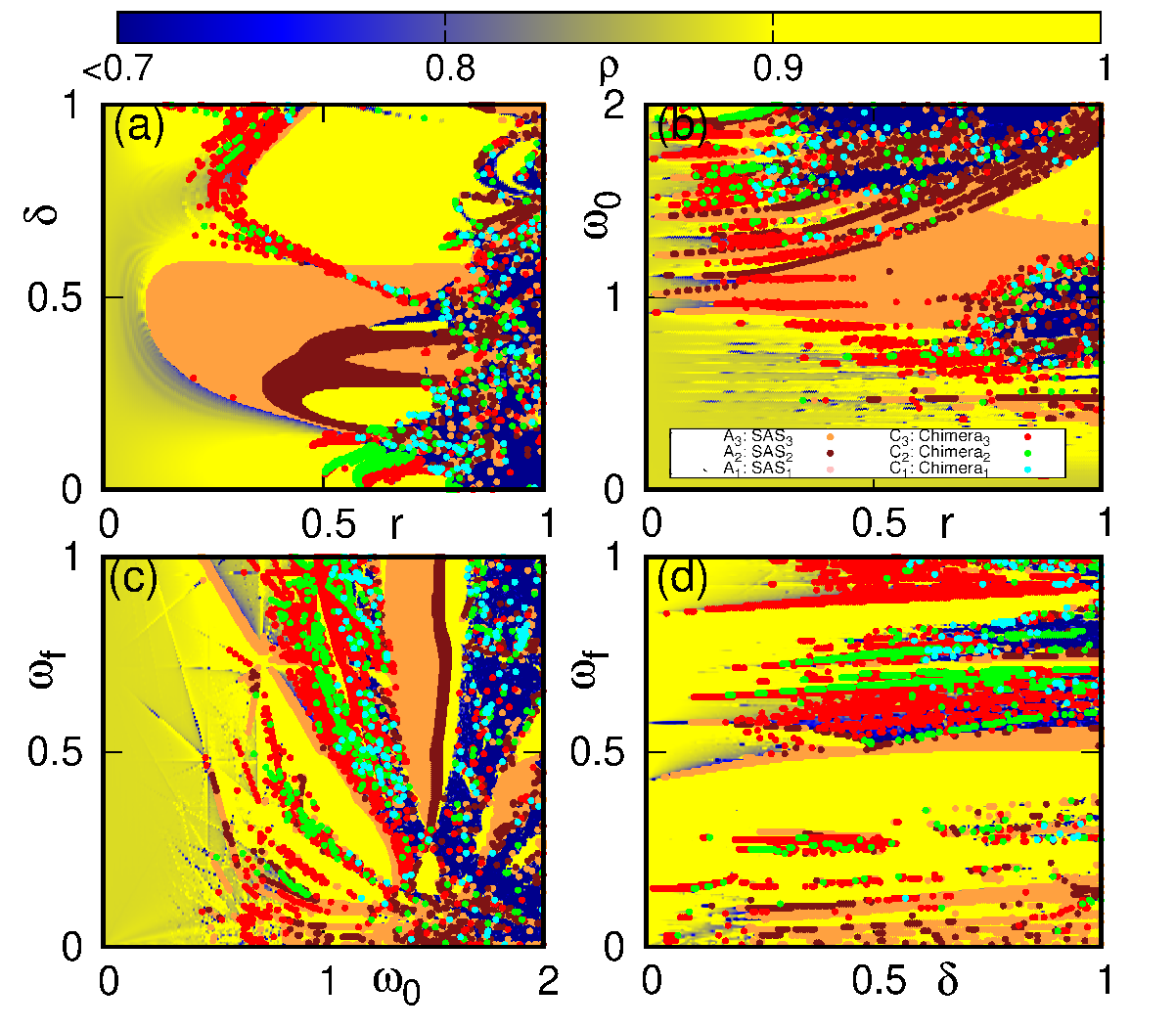}
    \caption{Dynamical regimes of the Van der Pol system under variation of frequency-modulation parameters. (a) $r$–$\delta$ plane ($\omega_0 = 1.0$, $\omega_f = 0.1$). (b) $r$–$\omega_0$ plane ($\delta = 0.5$, $\omega_f = 0.1$). (c) $\omega_0$–$\omega_f$ plane ($r = 0.4$, $\delta = 0.5$). (d) $\delta$–$\omega_f$ plane ($r = 0.4$, $\omega_0 = 1.0$). Colors and markers denote different dynamical states.}
    \label{vdp_other_pars}
\end{figure}
Figure~\ref{vdp_other_pars}(a) presents the dynamical regimes in the $r$--$\delta$ parameter space for fixed $\omega_0 = 1.0$ and $\omega_f = 0.1$. The color coding follows the same convention as in Fig.~\ref{vdp_two_par} (and Table-\ref{tab1}), allowing direct comparison with the previously discussed $\mu$--$r$ plane. Distinct regions corresponding to global synchrony, chimera states (C$_1$, C$_2$, C$_3$), and CS–Anti-S states (A$_1$, A$_2$, A$_3$) can be clearly identified.

To further explore the role of the carrier frequency, Fig.~\ref{vdp_other_pars}(b) shows the distribution of dynamical states in the $r$--$\omega_0$ plane for fixed $\delta = 0.5$ and $\omega_f = 0.1$. Similarly, Fig.~\ref{vdp_other_pars}(c) illustrates the behavior in the $\omega_0$--$\omega_f$ plane for $r = 0.4$ and $\delta = 0.5$, highlighting the combined influence of carrier and modulation frequencies. Finally, the $\delta$--$\omega_f$ parameter space is shown in Fig.~\ref{vdp_other_pars}(d) for $r = 0.4$ and $\omega_0 = 1.0$, revealing how the modulation depth and frequency jointly affect the system dynamics.

Across all parameter planes, a rich variety of dynamical regimes is observed, including complete synchronization, complete incoherence, multiple chimera configurations, and coexistence of synchronization with anti-synchronization. These results demonstrate that the collective behavior of the system is highly sensitive to the characteristics of the external modulation. In particular, the interplay between modulation amplitude, depth, and frequency provides an effective mechanism for selecting and controlling different dynamical states, even in the absence of direct coupling among the oscillators.

\subsection{Phase reduction of the forced system}
\label{subsec:phase}

To obtain analytical insight into the emergence of synchronization and chimera states, we employ the phase reduction framework \cite{Kurebayashi13Phase}. Although the full dynamics of the system are governed by nonlinear equations in a higher-dimensional phase space, the essential features of collective behavior can be captured by focusing on the evolution of oscillator phases. This approach allows us to reduce the dynamics to a lower-dimensional description, facilitating the analysis of phase locking and stability \cite{Ermentrout81nm, Kuramoto84chemical}.

We begin by expressing Eq.~\eqref{vdpeq} in a general form \cite{bhandary23stability}
\begin{equation}\label{vdpgeneq}
    \frac{d\mathbf{X}_i}{dt}=\mathbf{F}(\mathbf{X}_i)+\mu r\mathbf{I}(\mathbf{X}_i)p(t),
\end{equation}
where $\mathbf{X}_i = [x_i, y_i]^T$, $\mathbf{F}(\mathbf{X}_i) = [y_i, \mu(1 - x_i^2)y_i - \omega^2 x_i]^T$ represents the intrinsic dynamics, and $\mathbf{I}({X}_i) = [0, (1 - x_i^2)y_i]^T$ characterizes the influence of the external modulation. The forcing term is given by $p(t) = \sin\left(\omega_0\left(t - \frac{\delta}{\omega_f}\cos(\omega_f t)\right)\right)$.

In the absence of forcing, the system $\dot{\mathbf{X}}_i = \mathbf{F}(\mathbf{X}_i)$ admits a stable limit cycle $S$ with period $T$ and natural frequency $\omega = 2\pi/T$. The state of the oscillator on this limit cycle can be parametrized as $\mathbf{X}_0(\theta)$, where $\theta$ denotes the phase. We define a phase function $\Theta(X)$ such that $\theta_i(t) = \Theta(\mathbf{X}_i(t))$, and introduce the phase sensitivity function (phase response curve, PRC) $\mathbf{Z}(\theta)$, which quantifies the response of the phase to perturbations.

Assuming that the dynamics remain close to the limit cycle, we approximate $\mathbf{X}_i(t) \approx \mathbf{X}_0(\theta_i(t))$ and obtain the reduced phase equation
\begin{eqnarray}
    \frac{d\theta_i}{dt}&=&\frac{d}{dt}\Theta(\mathbf{X}_i(t))=\nabla_{\mathbf{X}_i}\Theta(\mathbf{X}_i)\rvert_{\mathbf{X}_i=\mathbf{X}_i(t)}\frac{d\mathbf{}X_i}{dt},\nonumber\\
    &\approx& \nabla_{\mathbf{X}_i}\Theta(\mathbf{X}_i)\rvert_{\mathbf{X}_i=\mathbf{X}_0(\theta_i(t))}\cdot\Big[\mathbf{F}(\mathbf{X}_i)\nonumber\\
    &&~~~~~~~~~~~~~~~~ ~~~~~~~+\mu r\mathbf{I}(\mathbf{X}_i)p(t)\Big],\nonumber\\
    &=&\omega+\mu r\mathbf{Z}(\theta_i)\cdot \mathbf{I}(\theta_i)p(t).
\end{eqnarray}
where the dot denotes the standard inner product.

\begin{figure}
    \centering
    \includegraphics[width=0.48\textwidth]{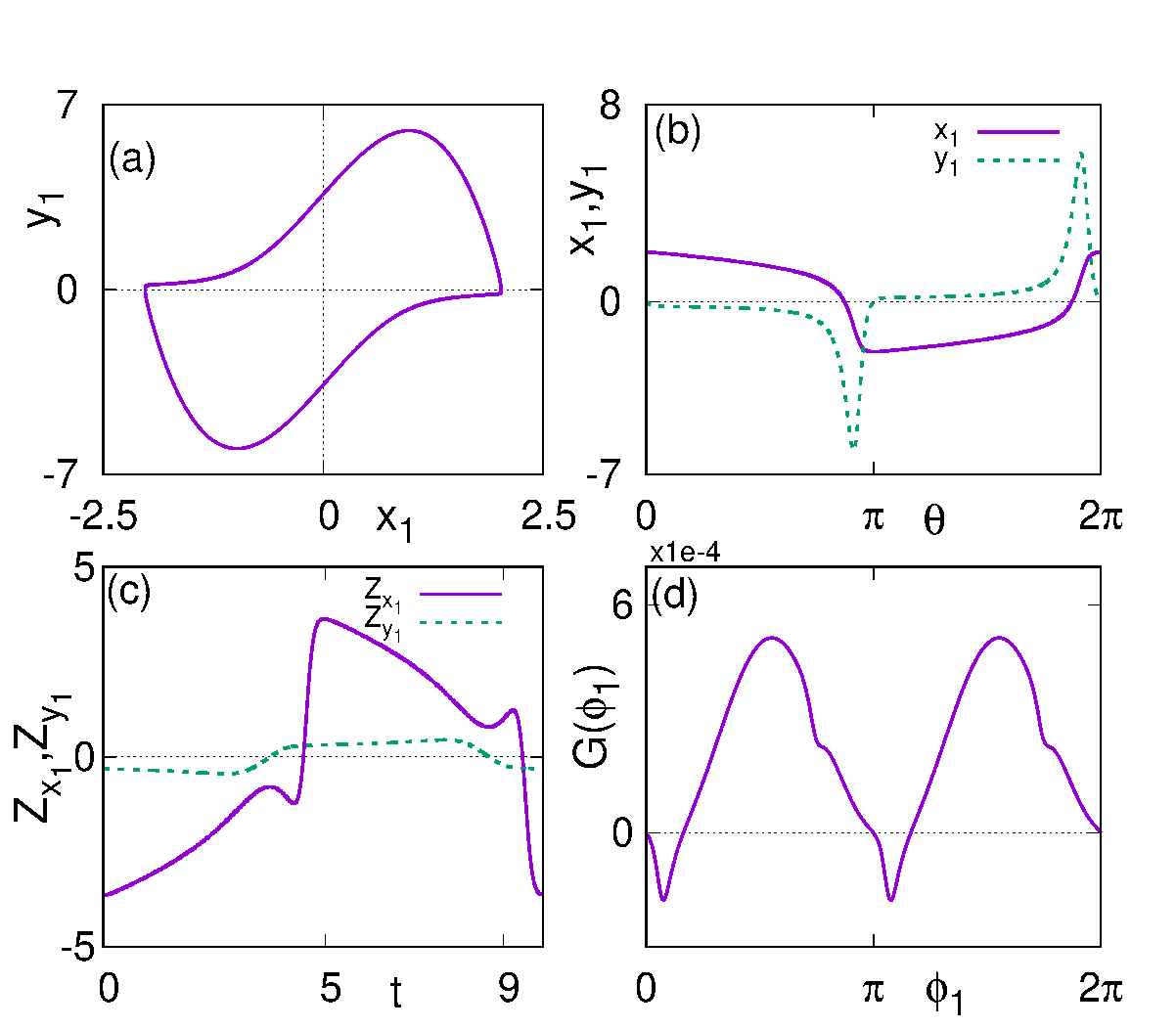}
    \caption{Phase reduction analysis of the Van der Pol system. (a) Limit cycle in phase space. (b) Corresponding time series. (c) Phase response curve $\mathbf{Z}(\theta)$ over one period. (d) Effective phase interaction function $G(\phi)$, where zeros indicate phase-locked states. Other parameters are: $\mu=3.75$, $r=0.2$ $\delta=0.5$, $\omega_0=1.0$, and $\omega_f=0.1$, and $N=3$.}
    \label{vdp_zg}
\end{figure}

The phase response curve is given by $\mathbf{Z}(\theta_i) = \nabla_{\mathbf{X}_i}\Theta(\mathbf{X}_i)\rvert_{\mathbf{X}_i = \mathbf{X}_0(\theta_i)}$, and describes how infinitesimal perturbations influence the phase evolution. Positive values of the PRC correspond to phase advancement, while negative values indicate phase delay.

\begin{figure}
    \centering
    \includegraphics[width=0.5\textwidth]{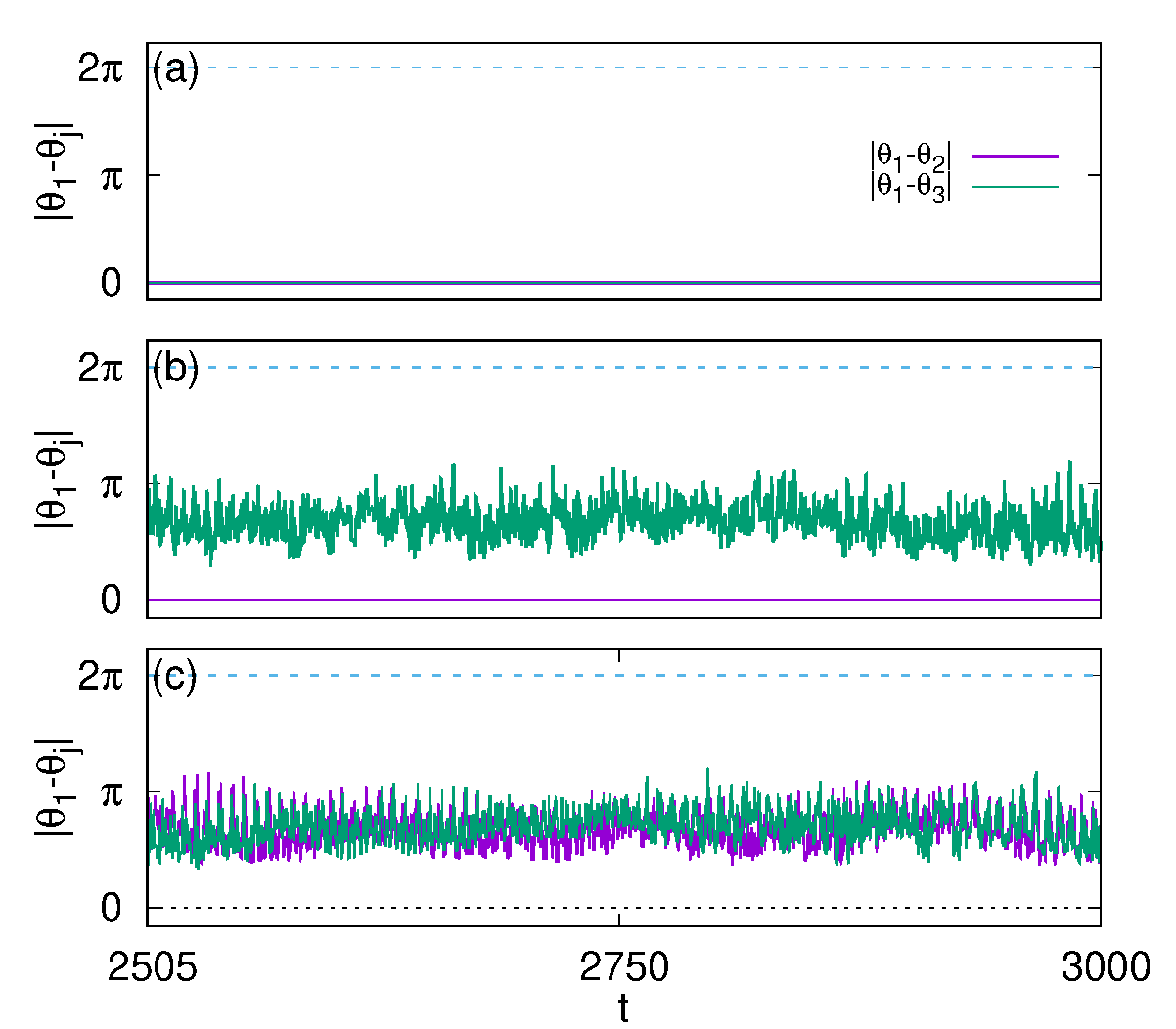}
    \caption{The averaged phase differences ($\lvert\phi_{1i}\rvert=\lvert\theta_{1}-\theta_{i}\rvert$) of the Van der Pol system for $10$ initial conditions around different dynamical zones using Eq.~\eqref{geq}. This illustrates different dynamical regimes: (a) global synchronization (decay to zero) for $\mu=3.75$, $r=0.2$, (b) minimal chimera state (partial locking) for $\mu=3.51$, $r=0.91$, and (c) incoherence (persistent fluctuations) for $\mu=4.06$, $r=0.95$. Other parameters are the same as Fig.~\ref{vdp_zg}.}
    \label{vdp_phase_diff}
\end{figure}

We compute $\mathbf{Z}(\theta)$ numerically using the adjoint method based on Malkin’s approach \cite{ermentrout2010mathematical, malkin1959some, bhandary23stability}
\begin{equation}\label{adjeq}
    \omega\frac{d\mathbf{Z}(\theta_i)}{d\theta_i}=-\mathbf{J}(\theta_i)^T \mathbf{Z}(\theta_i),
\end{equation}
subject to the normalization condition
\begin{equation}\label{normeq}
    \mathbf{Z}(\theta_i)\cdot\frac{d\mathbf{X}_0(\theta_i)}{d\theta_i}=1.
\end{equation}

We numerically calculate $Z(\theta)$ from Eqs.~\eqref{adjeq}-\eqref{normeq}. Thus, using this, we convert the van der Pol equation (Eq.~\eqref{vdpeqi}) into three-dimensional phase equations and study its phase dynamics. The determination of synchronous solutions is done by the calculations of the phase differences of the oscillators ($\phi_{1i}=\theta_1-\theta_i$, $i=2,3$) as follows:
\begin{eqnarray}\label{geq}
    \frac{d\phi_{1i}}{dt}&=&\frac{d\theta_1}{dt}-\frac{d\theta_i}{dt},\nonumber\\
    &=& \mu rp(t)[\mathbf{Z}(\theta_1)\mathbf{I}(\theta_1)-\mathbf{Z}(\theta_i)\mathbf{I}(\theta_i)],\nonumber\\
    &=& \mu rp(t)[Q(\theta_1)-Q(\theta_1-\phi_{1i})],
\end{eqnarray}
considering $Q(\theta)=\mathbf{Z}(\theta)\mathbf{I}(\theta)$ and $\theta_i=\theta_1-\phi_{1i}$. Eq.~\eqref{geq} still depends explicitly on the instantaneous phase and on the external forcing. To characterize the slow evolution of the phase difference, we introduce an effective phase interaction function by averaging the instantaneous phase-difference velocity over a sufficiently long time interval, given by
\begin{equation}\label{geq1}
    G(\phi)=\lim_{T_a\rightarrow\infty}\frac{\mu r}{T}\int_0^{T_a}p(t)\left[Q(\theta(t))-Q(\theta(t)-\phi)\right]dt.
\end{equation}
Here, $T_a$ is chosen sufficiently large compared to both the natural oscillation period and the characteristic time scale of the frequency modulation. Now, the averaged relative phase dynamics may be written as 
\begin{equation}\label{geqav}
    \frac{d\phi_{1i}}{dt}\simeq G(\phi_{1i}).
\end{equation}
The above equation provides a reduced description of the relative dynamics between oscillators in terms of phase difference. Although the oscillators are not directly coupled, the function $G(\phi)$ effectively governs their mutual interaction through the common external modulation. In this sense, $G(\phi)$ plays the role of an effective coupling function induced by the shared forcing.
Physically, it represents the instantaneous rate of change of the phase difference, and therefore determines whether oscillators tend to synchronize or drift apart. Fixed points of $G(\phi)$, defined by $G(\phi^*) = 0$, correspond to phase-locked states.

The stability of these states is determined by the sign of the local slope of $G(\phi)$ near $\phi^\ast$: stable synchronization occurs when perturbations decay, whereas instability leads to phase drift. Consequently, the coexistence of stable and drifting solutions of Eq.~(10) provides a natural mechanism for the emergence of chimera states, where some oscillators synchronize while others remain desynchronized.

In Fig.~\ref{vdp_zg}(a,b), we present the limit cycle and corresponding time series of the unforced system. The numerically computed PRC $\mathbf{Z}(\theta)$ is shown in Fig.~\ref{vdp_zg}(c), while Fig.~\ref{vdp_zg}(d) illustrates the corresponding $G(\phi)$ function. This indicates that the common modulation effectively induces a phase interaction among oscillators, mimicking a coupling mechanism that can stabilize clustered states and thus chimera.

Finally, the dynamical states of the system can be characterized through the phase differences $|\theta_1 - \theta_i|$. As shown in Fig.~\ref{vdp_phase_diff}, for $\mu = 3.75$ and $r = 0.2$, both phase differences decay to zero, indicating global synchronization. For $\mu = 3.51$ and $r = 0.91$, one phase difference converges to zero while the other remains fluctuating, corresponding to a minimal chimera state. In contrast, for $\mu = 4.06$ and $r = 0.95$, both phase differences exhibit persistent fluctuations, indicating global incoherence.

\section{Model-II: Time-delayed system with frequency modulated forcing}
\label{sec:ddehw}
Next, to demonstrate the generality and robustness of the proposed scheme, we extend the mechanism to a chaotic time-delay system \cite{Banerjee2012design} given by
\begin{equation}\label{ddehweq}
    \dot{x}_i=-ax_i-b(1+r\sin(\vartheta_0))f(x_{i_\tau}),
\end{equation}
where $i=1,2,3$, $f(x_\tau)\equiv f(x(t-\tau))=-0.5n(|x(t-\tau)|+x(t-\tau))+m\tanh(lx(t-\tau))$, $a\in\mathcal{R}^+$, $b\in\mathcal{R}^+$ are the system parameters and $\{n,m,l\}\in\mathcal{R}^+$ are the parameters controlling the nonlinearity of the system. Here also the $\vartheta_0$ follows the Eq.~\eqref{phieq}. Henc, the frequency-modulated parameter becomes $b(t)=b\left[1 + r\sin\left\{\omega_0\left(t - \frac{\delta}{\omega_f}\cos(\omega_f t)\right)\right\}\right]$.

\begin{figure}
    \centering
    \includegraphics[width=0.48\textwidth]{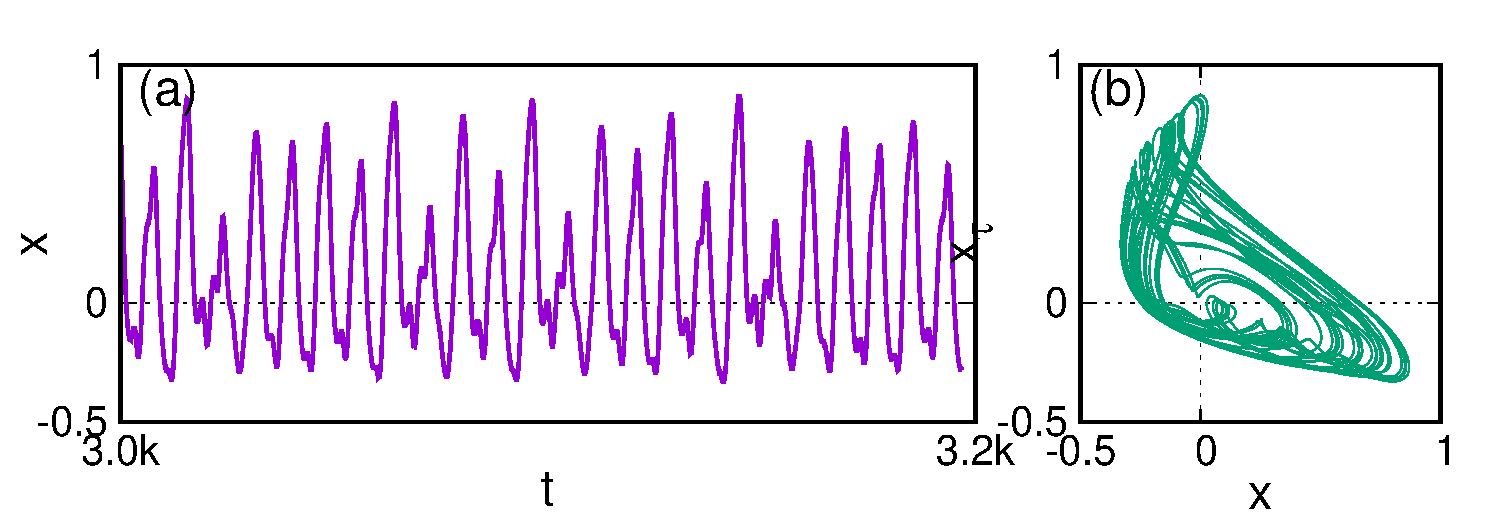}
    \caption{Dynamics of the time-delayed system of Eq.~\eqref{ddehweq} without forcing. (a) Chaotic time series. (b) Phase portrait in the $x$–$x_\tau$ plane. Parameters: $a = 1$, $b = 1.7$, $\tau = 3.2$, $n = 1.15$, $m = 0.97$, and $l = 2.19$.}
    \label{ddehw_ts}
\end{figure}

The system emerges chaos for some range of proper parameters and delay (details in Ref.~\cite{Banerjee2012design}). In Fig.~\ref{ddehw_ts}(a), we show the chaotic time series and the associated phase plane in $x-x_\tau$ space for the autonomous system ($r=0$) in Fig.~\ref{ddehw_ts}(b) for $a=1$, $b=1.7$, $\tau=3.2$, $n=n=1.15$, $m=0.97$, $l=2.19$.

\begin{figure}
    \centering
    \includegraphics[width=0.48\textwidth]{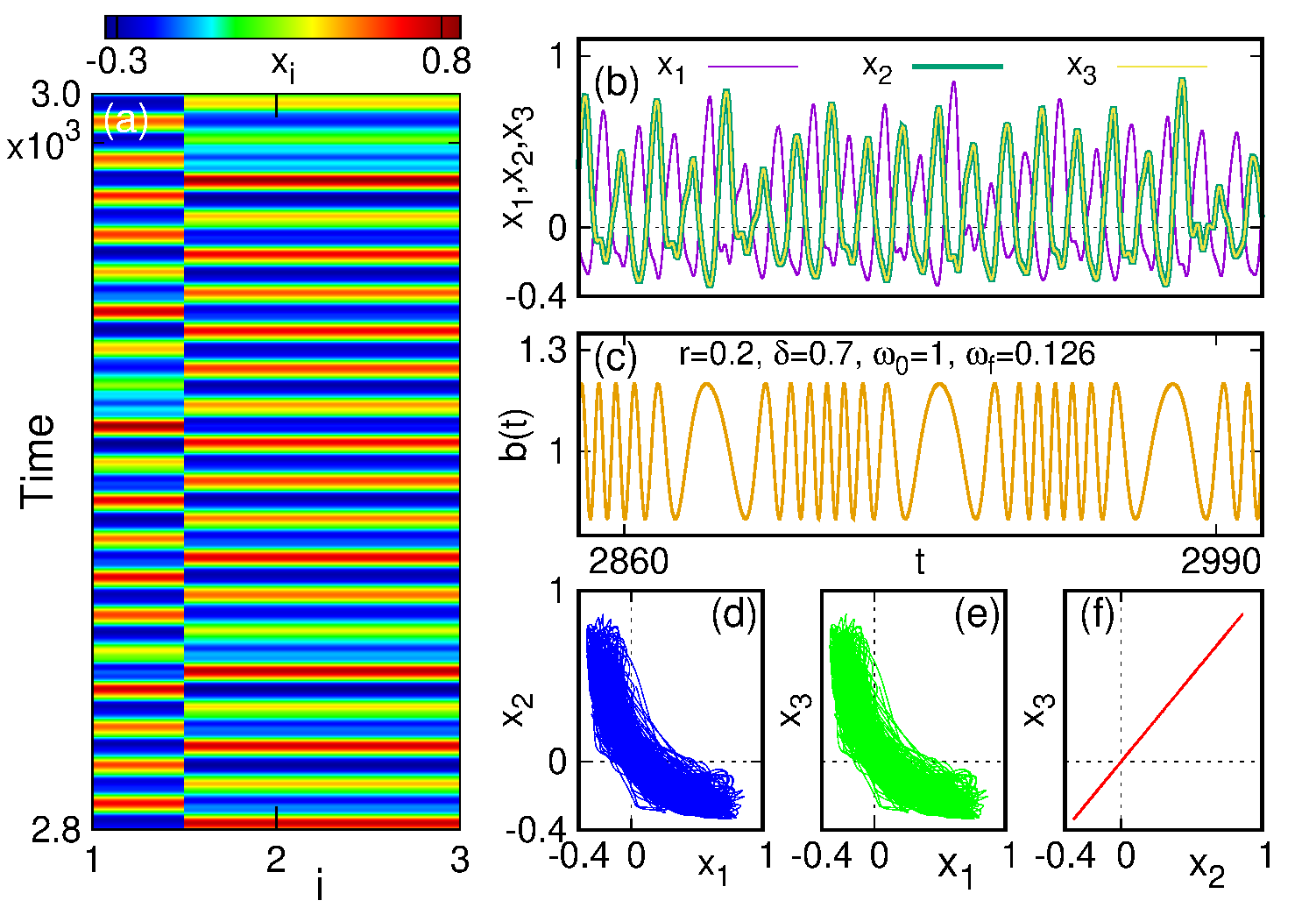}
    \caption{Minimal chimera in the time-delayed system. (a) Spatiotemporal evolution. (b) Time series. (c) Frequency-modulated forcing $b(t)$. (d),(e) Phase portraits in $x_1$–$x_2$ and $x_1$–$x_3$ showing desynchronization. (f) Phase portrait in $x_2$–$x_3$ showing synchronization. Parameters: $b = 1.47$, $\tau = 3.0$, $r = 0.2$, $\delta = 0.7$, $\omega_0 = 1.0$, $\omega_f = 0.126$.}
    \label{ddehw_fm_sptmp}
\end{figure}

\subsection{Results: Minimal chimera in a network of
frequency modulated parameter driven time delayed
oscillators}
\label{subsec:restd}

Chimera states have been extensively investigated in limit-cycle and phase oscillator systems \cite{kuramoto2002coexistence, panaggio2015chimera}, whereas their realization in chaotic time-delayed systems remains comparatively less explored. In this context, we demonstrate that a network of three uncoupled time-delayed oscillators can exhibit minimal chimera states when subjected to a common frequency-modulated forcing in a parameter.

\begin{figure*}
    \centering
    \includegraphics[width=0.8\textwidth]{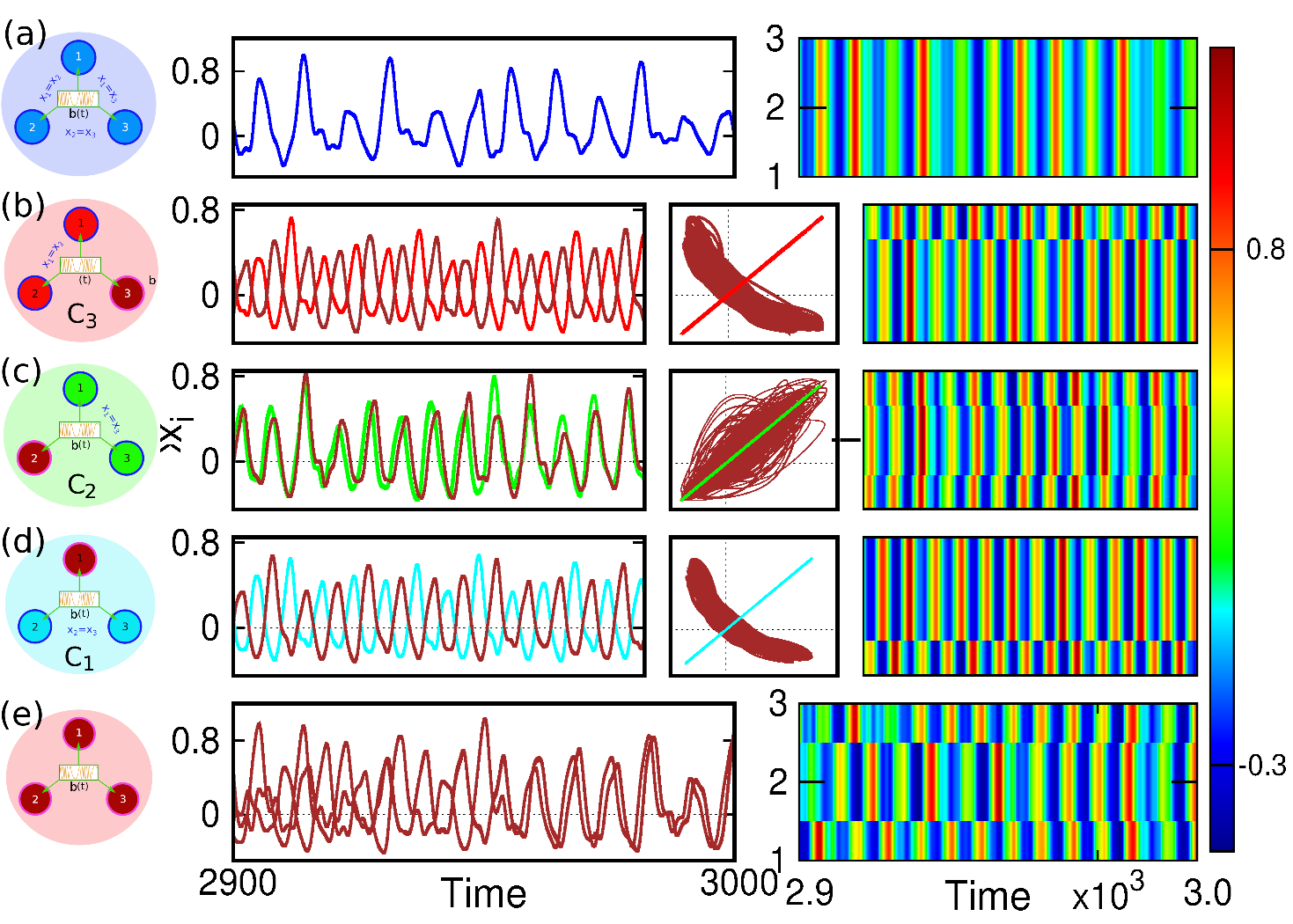}
    \caption{Representative dynamical states of the time-delayed system. (a) Global synchronization. (b) Chimera state C$_3$. (c) Chimera state C$_2$. (d) Chimera state C$_1$. (e) Global incoherence. Left column: time series; middle: phase-space projections; right: spatiotemporal plots. Parameters are given in the text.}
\label{ddehw_sync_usync}
\end{figure*}

In the absence of modulation ($r = 0$), the oscillators evolve independently, and no synchronization is observed. When the modulation is introduced ($r \neq 0$), the system becomes nonautonomous, and the common forcing begins to influence the dynamics of all oscillators simultaneously. Under suitable parameter conditions, this leads to the emergence of partial synchronization, where a subset of oscillators lock together while others remain incoherent, hence representing the chimera state.

A representative example of such behavior is shown in Fig.~\ref{ddehw_fm_sptmp} for $b = 1.47$, $\tau = 3.0$, $r = 0.2$, $\delta = 0.7$, $\omega_0 = 1.0$, and $\omega_f = 0.126$. The spatiotemporal plot in Fig.~\ref{ddehw_fm_sptmp}(a) indicates that $x_2$ and $x_3$ evolve coherently, while $x_1$ remains desynchronized (C$_1$). This observation is consistent with the corresponding time series shown in Fig.~\ref{ddehw_fm_sptmp}(b), where the trajectories of $x_2$ and $x_3$ overlap, in contrast to $x_1$. The temporal evolution of $b(t)$ is shown in Fig.~11(c).

Further confirmation is obtained from the phase-space projections. The $x_1$--$x_2$ and $x_1$--$x_3$ projections [Fig.~\ref{ddehw_fm_sptmp}(d,e)] show scattered distributions, indicating the absence of synchronization involving $x_1$. In contrast, the $x_2$--$x_3$ projection [Fig.~\ref{ddehw_fm_sptmp}(f)] collapses onto a diagonal line, demonstrating complete synchronization between these two oscillators. These results establish the presence of a minimal chimera state in the time-delayed system.

The observed behavior can be interpreted in terms of an effective interaction induced by the common external modulation. Although the oscillators are not directly coupled, the shared frequency-modulated forcing acts as a global drive. Due to differences in initial conditions and the inherent nonlinear response of the system, the oscillators do not respond identically to this driving. As a result, some oscillators lock to the modulation and synchronize, while others remain incoherent, leading to the formation of clustered states.

To explore the robustness of these dynamical regimes, we examine the system behavior over a range of parameter values by varying $b$ and $r$, while fixing $\delta = 0.4$, $\omega_0 = 1.0$, and $\omega_f = 0.126$. The results are summarized in Fig.~\ref{ddehw_sync_usync}, where the left column shows schematic representations, the middle column presents time series, and the right column displays the corresponding spatiotemporal patterns.

For $b = 1.3$ and $r = 0.4$, the system exhibits global synchronization [Fig.~\ref{ddehw_sync_usync}(a)], with all oscillators evolving identically. Increasing $b$ and $r$ leads to the emergence of chimera states. For instance, Fig.~\ref{ddehw_sync_usync}(b) shows the C$_3$ configuration ($b = 1.6$, $r = 0.58$), where $x_1$ and $x_2$ are synchronized while $x_3$ remains incoherent. Similarly, the C$_2$ state is observed in Fig.~\ref{ddehw_sync_usync}(c) for $b = 1.375$ and $r = 0.37$, where $x_1$ and $x_3$ synchronize and $x_2$ remains desynchronized. The C$_1$ configuration, shown in Fig.~\ref{ddehw_sync_usync}(d) for $b = 1.3$ and $r = 0.3$, corresponds to synchronization between $x_2$ and $x_3$, with $x_1$ remaining incoherent. Finally, for $b = 1.65$ and $r = 0.3$, the system exhibits global incoherence [Fig.~\ref{ddehw_sync_usync}(e)], where none of the oscillators synchronize.

These results demonstrate that frequency-modulated forcing can induce a wide range of collective behaviors in time-delayed systems, including global synchrony, minimal chimera, and complete incoherence. The transitions between these states are governed by the interplay between the system parameters and the characteristics of the external modulation, highlighting the role of common forcing as an effective mechanism for organizing collective dynamics in uncoupled systems.

It should be noted that no anti-synchronized state was detected for the time-delayed model within the parameter ranges investigated here. This absence is specific to the explored parameter domain and should not be interpreted as a general prohibition of anti-synchronization in delayed dynamical systems. In contrast, anti-synchronization is clearly observed for the frequency-modulated van der Pol oscillators and is therefore reported separately in the corresponding state classification previously.

\subsection{Dynamics in $b-r$ space}
\label{sub:br}

The results presented in Fig.~\ref{ddehw_sync_usync} already indicate that the system exhibits a wide range of dynamical behaviors as the parameters are varied. To examine this systematically, we now explore the structure of the $b$--$r$ parameter space for different values of the modulation depth $\delta$. As initial conditions, we use: $x_1(0)=0.9$, $x_2(0)=0.7$, and $x_3(0)=0.5$, and the same for the history throughout the investigation.
\begin{figure}
    \centering
    \includegraphics[width=0.48\textwidth]{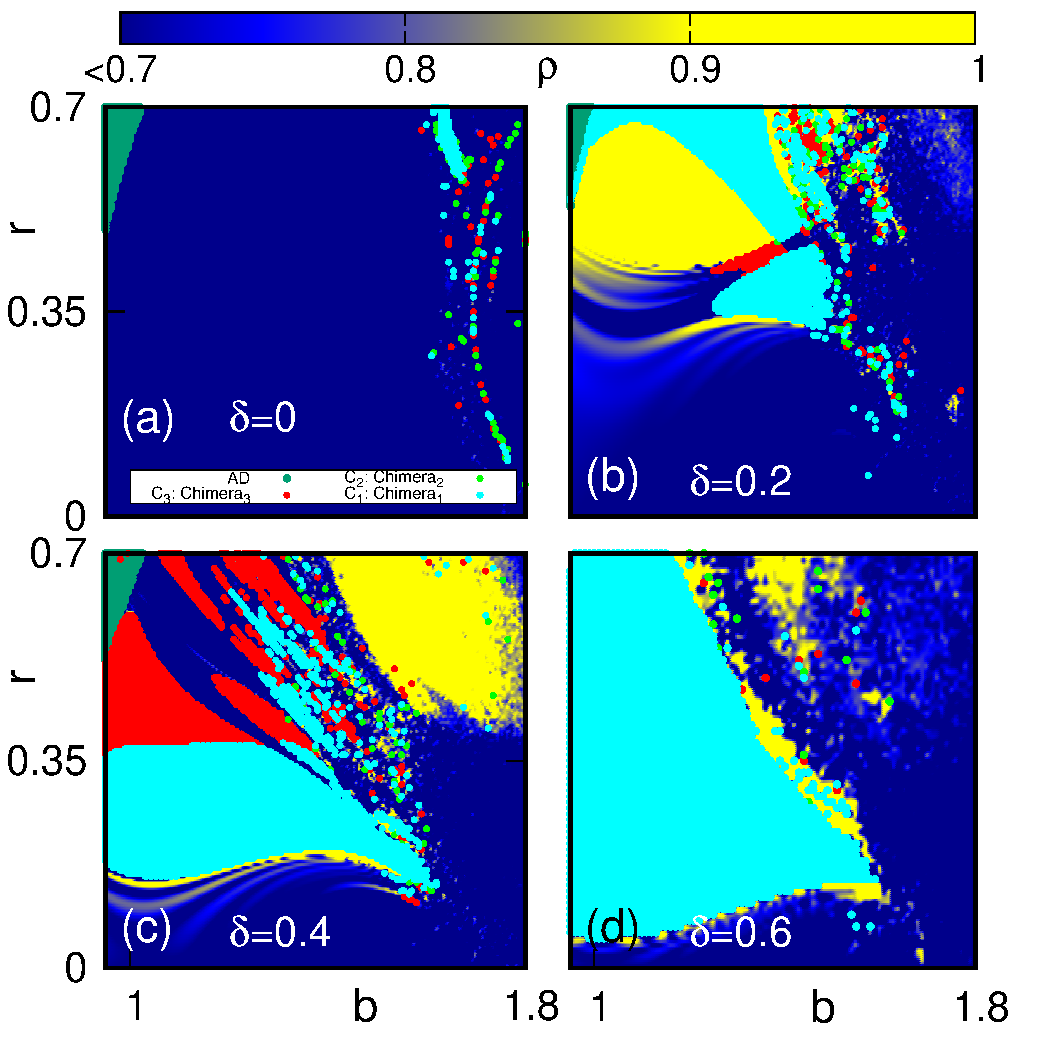}
    \caption{Phase diagrams of the time-delayed system in the $b$–$r$ plane for (a) $\delta = 0$, (b) $\delta = 0.2$, (c) $\delta = 0.4$, and (d) $\delta = 0.6$. Colors denote dynamical regimes, including amplitude death, synchronization, and chimera states.}
    \label{ddehw_b_eta}
\end{figure}

The corresponding phase diagrams are shown in Fig.~\ref{ddehw_b_eta}. For $\delta = 0$ [Fig.~\ref{ddehw_b_eta}(a)], the system predominantly remains in an unsynchronized state across most of the parameter space. However, in a region characterized by relatively small $b$ and larger $r$, the system settles into amplitude death (AD), indicated by the dark green region in the figure.

As the modulation depth is increased, the dynamical landscape undergoes a significant reorganization. For $\delta = 0.2$ [Fig.~\ref{ddehw_b_eta}(b)], regions corresponding to global synchronization, chimera states, and incoherent dynamics begin to emerge. The chimera configurations are classified as C$_3$, C$_2$, and C$_1$, denoted by red, green, and cyan markers, respectively, following the same convention as in Table-\ref{tab1}. Notably, in contrast to the previous model, no CS–Anti-S (A$_i$) states are observed in the time-delayed system.

Further increasing the modulation depth enhances the prevalence of chimera states. For $\delta = 0.4$ [Fig.~\ref{ddehw_b_eta}(c)], the regions supporting chimera configurations expand noticeably within the parameter space. This trend continues for $\delta = 0.6$ [Fig.~\ref{ddehw_b_eta}(d)], where chimera states occupy a substantial portion of the diagram, while the amplitude death region disappears entirely.

Overall, these results demonstrate that the modulation depth $\delta$ plays a crucial role in shaping the collective dynamics of the system. In particular, increasing $\delta$ promotes the emergence and stabilization of chimera states, while suppressing trivial or inactive regimes such as amplitude death. This behavior further supports the interpretation that frequency-modulated forcing acts as an effective organizing mechanism, capable of inducing structured collective dynamics in the absence of direct coupling.

\begin{figure}
    \centering
    \includegraphics[width=0.48\textwidth]{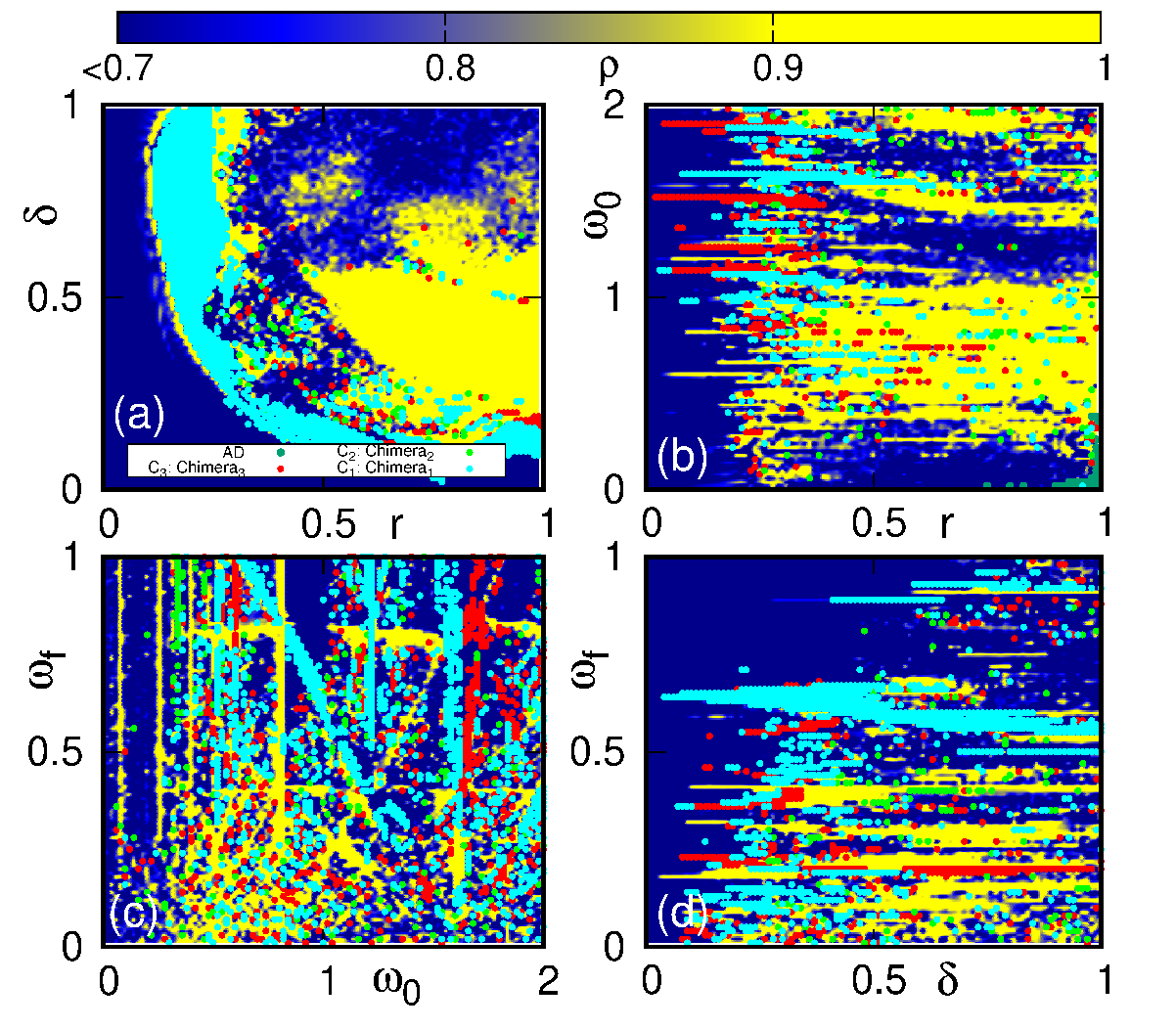}
    \caption{Order parameter $\rho$ for the time-delayed system in different parameter planes: (a) $r$–$\delta$, (b) $r$–$\omega_0$, (c) $\omega_0$–$\omega_f$, and (d) $\delta$–$\omega_f$. Parameters: $b = 1.47$, $\tau = 3.0$.}
    \label{ddehw_other_pars}
\end{figure}

\subsection{Effect of variation of Frequency Modulation
parameters for the time-delayed system}
\label{subsec:fmparstd}

We now examine how the collective dynamics of the time-delayed system depend on the parameters governing the frequency-modulated driving. In particular, we focus on the roles of $r$, $\delta$, $\omega_0$, and $\omega_f$, which together determine the amplitude, depth, and temporal structure of the modulation.

We begin with the $r$--$\delta$ parameter plane for fixed $\omega_0 = 1.0$ and $\omega_f = 0.126$, as shown in Fig.~\ref{ddehw_other_pars}(a). The diagram reveals a clear organization of dynamical regimes, including synchronized, incoherent, and chimera states, identified using the same color scheme as Table-\ref{tab1}. Notably, chimera states are predominantly observed for relatively small values of both $r$ and $\delta$, indicating that moderate modulation is sufficient to induce partial synchronization in the system.

Next, we consider the influence of the carrier frequency by exploring the $r$--$\omega_0$ parameter space [Fig.~\ref{ddehw_other_pars}(b)] for $\delta = 0.5$ and $\omega_f = 0.126$. Variations in $\omega_0$ lead to noticeable changes in the distribution of dynamical states, highlighting its role in shaping the response of the system to external driving.

The combined effect of carrier and modulation frequencies is illustrated in the $\omega_0$--$\omega_f$ plane [Fig.~\ref{ddehw_other_pars}(c)] for $r = 0.4$ and $\delta = 0.4$. This representation emphasizes how the interplay between these two frequencies influences the emergence of synchronization and chimera states.

Finally, Fig.~\ref{ddehw_other_pars}(d) presents the $\delta$--$\omega_f$ parameter space for $r = 0.4$ and $\omega_0 = 1.0$. The results show that both the modulation depth and frequency significantly affect the stability and distribution of the observed dynamical regimes.

Overall, these parameter-space explorations demonstrate that the characteristics of the frequency-modulated forcing provide an effective means of controlling the collective behavior of the system. Even in the absence of direct coupling, appropriate tuning of the modulation parameters can selectively promote synchronization, sustain chimera states, or drive the system toward incoherence.

\section{Electronic Experiment}
\label{sec:expt}

\subsection{Electronic circuit realization}
\label{subsec:ckt}

\begin{figure}
    \centering
    \includegraphics[width=0.45\textwidth]{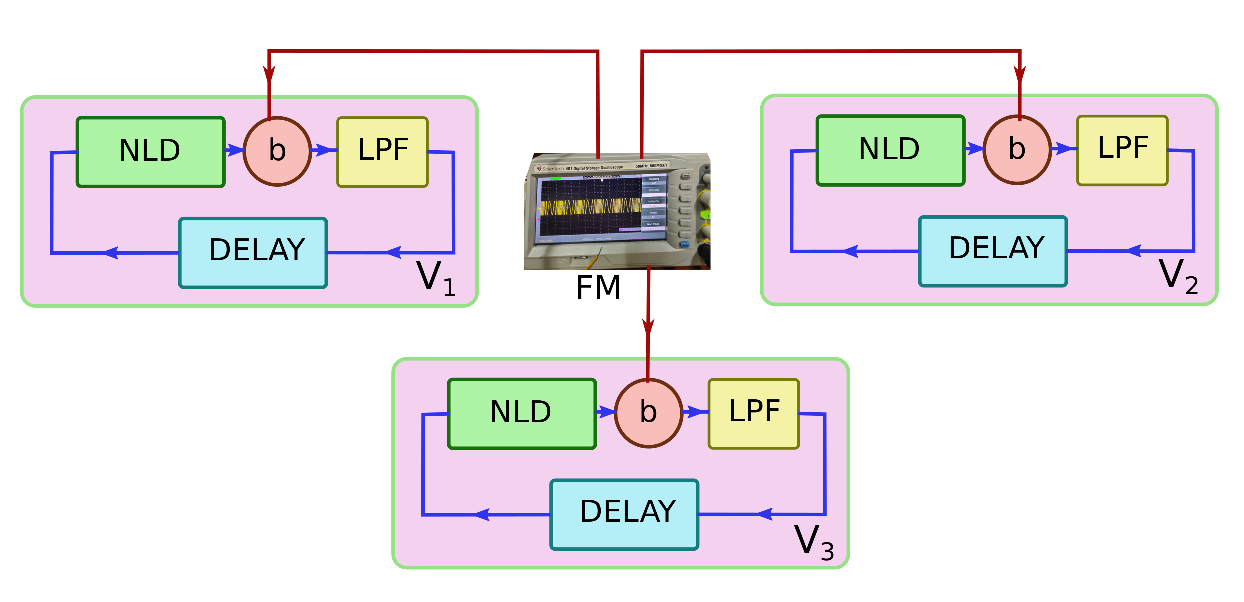}
    \caption{Schematic diagram of the electronic circuit used for experimental realization of the time-delayed system.}
    \label{ckt_schema}
\end{figure}

To validate the proposed time-delayed system in a physical setting, we implement the network of time-delayed systems using an analog electronic circuit assembled on a breadboard. A schematic representation of the setup is shown in Fig.~\ref{ckt_schema}.

\begin{figure}
    \centering
    \includegraphics[width=0.45\textwidth]{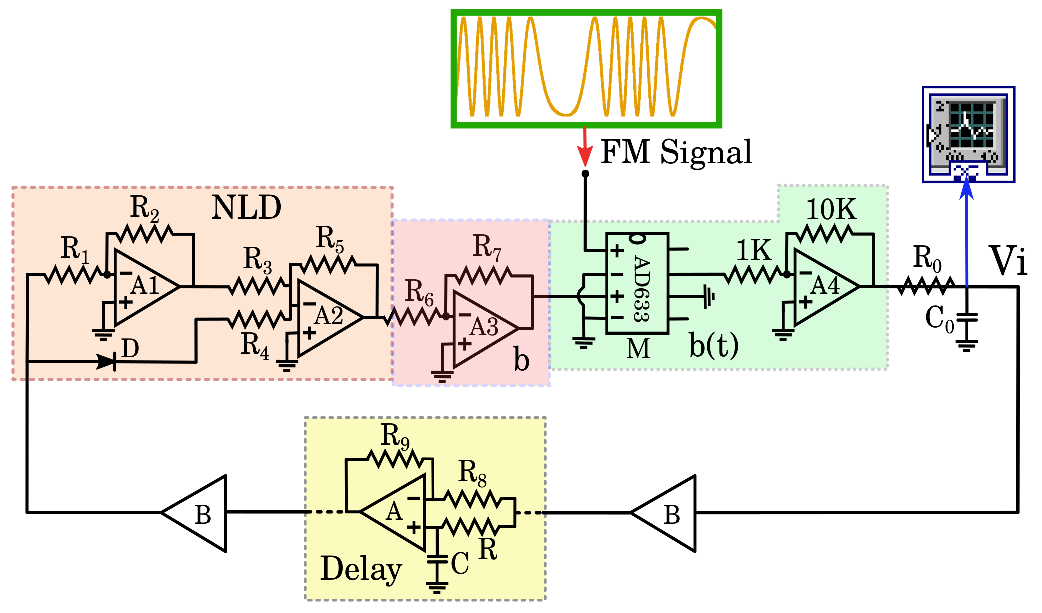}
    \caption{Detailed circuit implementation of the time-delayed oscillator, including nonlinear device (NLD), multiplier, delay units, and filtering components.}
    \label{ckt_fm}
\end{figure}

The detailed circuit configuration is presented in Fig.~\ref{ckt_fm}. The implementation employs TL074 (quad JFET operational amplifiers) along with AD633JN multiplier chips, powered by a $\pm 14$ V supply.

All passive components are selected with a tolerance of $\pm 5\%$. The nonlinear device (NLD) is realized using a 1N4148 diode in combination with resistors $R_1=10$ k\ohm, $R_2=18$ k\ohm, $R_3=18$ k\ohm, $R_4=6.5$ k\ohm, and $R_5=10$ k\ohm, $R_6=R_7=10$ k\ohm. The frequency-modulated gain $b(t)$ is implemented through the AD633JN multiplier. Specifically, the output of amplifier A3 is connected to the non-inverting input (pin 3), while the non-inverting input (pin 1) receives the modulation signal (FM Signal) generated by a function generator (GW-Instek MFG-2230M). The low-pass filter (LPF) section consists of $R_0=1$ k\ohm.

The time-delay element is realized using an all-pass filter (APF) configuration with $R_8=R_9=2.2$ k\ohm~ and $R=10$ k\ohm, along with a capacitor $C=10$ nF \cite{biswas2018book}. This arrangement introduces a delay of approximately $T_D \approx RC = 0.1$ ms, corresponding to a dimensionless delay $\tau = \frac{RC}{R_0 C_0} = 1$. To achieve $\tau = 3$, three such APF units are connected in series. The delay can be tuned continuously by adjusting the resistance $R$. Buffer stages (B) are included to ensure proper impedance matching throughout the circuit.

Let $V_i(t)$ denote the voltage across the capacitor $C_0$ in the LPF section. The circuit dynamics can then be expressed as
\begin{equation}\label{ckteq}
    R_0C_0\frac{dV_i(t)}{dt}=-V_i(t)-\frac{R_7}{R_6}\times \mbox{(FM Signal)}\times f\big(V_i(t-T_D)\big),
\end{equation}
where $f(\cdot)$ is given by
\begin{equation}\label{cktnleq}
\begin{split}
    f\big(V_i(t&-T_D)\big)=-\frac{R_4}{R_5}\big(\lvert V_i(t-T_D)\rvert+V_i(t-T_D)\big)\\
    &+\frac{R_5}{R_3}\beta V_{sat}\tanh\left(w\frac{R_2}{R_1}\frac{V_i(t-T_D)}{V_{sat}}\right).
\end{split}
\end{equation}
The first term in the right hand side originates from the diode-based nonlinear element, while the second arises from the amplifier A1 \cite{Abdelfattah2006Modeling}. The gain $b(t)$ is externally imposed through a frequency-modulated signal of the form
\begin{equation}\label{beqexpt}
    b(t)\equiv b\times \mbox{(FM Signal)}=b\Big(1+r\sin\big(\omega_0(t-\frac{\delta}{\omega_f}\cos(\omega_f t))\big)\Big),
\end{equation}
that gives the frequency-modulated parameter and is provided by the function generator. 

To relate the circuit to the dimensionless model, we introduce the transformations $t = \frac{t}{R_0C_0}$, $\tau = \frac{T_D}{R_0C_0}$, $x_i = \frac{V_i(t)}{V_{sat}}$, and $x_i(t-\tau) = \frac{V_i(t-T_D)}{V_{sat}}$, along with parameter rescaling $b=\frac{R_7}{R_6}$, $n_1 = \frac{R_5}{R_4}$, $m_1 = \beta\frac{R_5}{R_3}$, and $l_1 = w\frac{R_2}{R_1}$. Under these transformations, Eq.~\eqref{ckteq} reduces to
\begin{equation}\label{ckteqdimless}
    \frac{dx_i}{dt}=-x_i(t)-b(t)f\big(x_i(t-\tau)\big),
\end{equation}
with the nonlinear function
\begin{equation}\label{cktnleqdimless}
    \begin{split}
        f(x_{i_\tau}) = -0.5n_1\big(\lvert x_{i_\tau}\rvert + x_{i_\tau}\big) + m_1\tanh(l_1 x_{i_\tau}).
    \end{split}
\end{equation}
Thus, the circuit equations recover the form of the theoretical model [Eq.~\eqref{ddehweq}] with $a=1$, demonstrating that the electronic implementation faithfully reproduces the dynamics of the time-delayed system for appropriate choices of $(n_1, m_1, l_1)$.

\subsection{Experimental Results}
\label{subsec:cexptresult}

\begin{figure}
    \centering
    \includegraphics[width=0.48\textwidth]{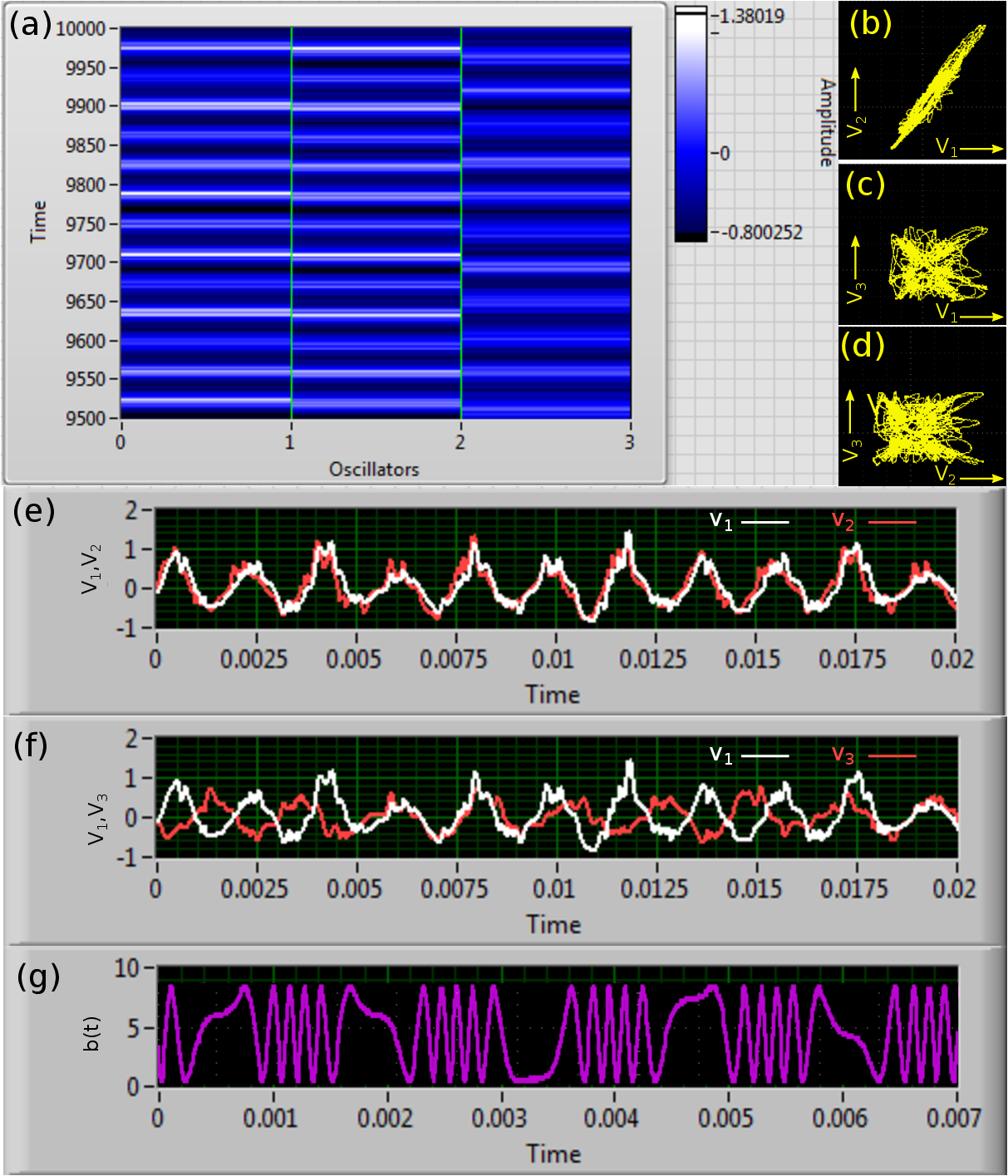}
    \caption{Experimental observation of minimal chimera. (a) Spatiotemporal evolution of voltages. (b) Phase portrait $V_1$–$V_2$ showing synchronization. (c),(d) Phase portraits $V_1$–$V_3$ and $V_2$–$V_3$ showing desynchronization. (e) Time series of $V_1$ and $V_2$. (f) Time series of $V_1$ and $V_3$. (g) Frequency-modulated input signal $b(t)$.}
    \label{expt_daq}
\end{figure}

The experimental data are acquired using a National Instruments Data Acquisition system (DAQ NI USB-6351, 8 inputs, 2 outputs, maximum sampling rate 1.25 MS/s) interfaced with a computer through LabVIEW environment \cite{labview2014}. The corresponding data acquisition and visualization workflow is illustrated in Appendix-\ref{app:dac}. In this setup, signals are recorded via the DAQ Assistant module, while time series and spatiotemporal representations are generated using the Waveform Graph and Intensity Graph tools, respectively.

The frequency-modulated parameter $b(t)$ is generated externally using a function generator (GW-Instek MFG-2230M) and applied into the circuit through multiplier units to realize the desired modulation. The experimental control parameters are chosen as follows: amplitude $= 4.0$ V (corresponding to $b$), DC offset $= 2.2$ V (corresponding to $br$), FM deviation $= 4$ kHz (representing $\omega_0$), and FM frequency $= 765$ Hz (representing $\omega_f$). These settings reproduce the frequency-modulated signal defined in Eq.~\eqref{beqexpt}.

The experimental observations are summarized in Fig.~\ref{expt_daq}. The spatiotemporal plot in Fig.~\ref{expt_daq}(a) shows the temporal evolution of the voltages across the three oscillators, where the amplitude is encoded in color. A clear separation of dynamical behavior is evident: $V_1$ and $V_2$ evolve coherently, while $V_3$ remains distinct, indicating partial synchronization and hence chimera. This behavior is further confirmed through phase-space projections. The $V_1$--$V_2$ plane [Fig.~\ref{expt_daq}(b)] exhibits a narrow band around the diagonal, consistent with complete synchronization. The small spread around the ideal line reflects unavoidable experimental imperfections, including parameter mismatch, intrinsic noise, and hardware limitations. In contrast, the projections in the $V_1$--$V_3$ and $V_2$--$V_3$ planes [Fig.~\ref{expt_daq}(c,d)] show dispersed trajectories, indicating the absence of synchronization involving $V_3$.

The time-domain signals provide additional evidence. As shown in Fig.~\ref{expt_daq}(e), the traces of $V_1$ and $V_2$ overlap almost completely, confirming their synchronized evolution. On the other hand, Fig.~\ref{expt_daq}(f) shows that $V_1$ and $V_3$ evolve independently, consistent with desynchronized behavior. The corresponding frequency-modulated input signal is displayed in Fig.~\ref{expt_daq}(g).

Taken together, these observations demonstrate that two oscillators ($V_1$, $V_2$) synchronize while the third ($V_3$) remains incoherent, which is the defining signature of a minimal chimera state (C$_3$ in this case). The experimental results are in good agreement with the numerical findings (cf. Fig.~\ref{ddehw_sync_usync}(b)), thereby confirming that frequency-modulated driving alone can induce chimera states in the absence of direct coupling. We note that other chimera configurations have also been observed experimentally under different parameter settings, although they are not presented here for brevity.

The experimental observation of the minimal chimera states also provides an indication of their robustness against realistic perturbations. Unlike the ideal deterministic numerical model, an electronic circuit is inevitably affected by thermal fluctuations, component tolerances, external disturbances, and measurement noise. The persistence of the predicted dynamical states in the experiment therefore suggests that the observed minimal chimeras are not confined to an ideal noise-free system. This observation should, however, be understood as robustness against inherent experimental fluctuations, rather than persistence under arbitrarily strong stochastic forcing.

\section{Conclusions}
\label{sec:conc}
In this work, we have demonstrated that minimal chimera states can arise in a system of three completely uncoupled oscillators when subjected to a common frequency-modulated driving. This finding challenges the conventional understanding that coupling among oscillators is a necessary ingredient for the emergence of chimera states. Instead, our results show that an appropriately structured external modulation can effectively induce a coexistence of coherent and incoherent dynamics, even in the absence of direct interactions. The results suggest that common external modulation can act as an effective coupling mechanism, enabling clustering and chimera formation without direct interactions.

Through systematic numerical analysis, we have identified distinct dynamical regimes, including global synchrony, global incoherence, and minimal chimera, arising from the interplay between intrinsic system parameters and the characteristics of the modulation. The use of maximal Lyapunov exponents provided insight into the stability of synchronized solutions, while the synchronization order parameter enabled a clear distinction between different collective states. Furthermore, the phase reduction approach offered a complementary analytical perspective, revealing how modulation-induced phase dynamics govern the onset of synchronization and desynchronization.

Importantly, the proposed mechanism is not restricted to a specific class of systems. By extending the analysis to a time-delayed and non-delayed (R\"{o}ssler system in Appendix-A) chaotic oscillators, we have shown that the emergence of minimal chimera under frequency-modulated forcing persists beyond simple limit-cycle dynamics. The experimental realization using an electronic circuit further substantiates the robustness of the phenomenon, demonstrating that such states are not merely numerical artifacts but can be observed in real physical systems despite unavoidable imperfections. This constitutes one of the first experimental demonstrations of chimera states arising purely from external modulation without coupling.

From a broader perspective, these results suggest that common external driving can act as an effective organizing factor in nonlinear systems, capable of inducing structured collective behavior without explicit coupling. This opens up new directions for controlling and engineering chimera-like states in systems where direct interactions are weak, absent, or difficult to manipulate. Potential implications may extend to fields such as neuroscience, where shared inputs influence neuronal populations, as well as to engineered systems involving distributed oscillators and signal-driven synchronization.

Overall, the present study establishes frequency-modulated driving as a viable and general route to chimera formation in uncoupled systems, thereby enriching the current understanding of cooperative dynamics in driven nonlinear media.

\section*{Data Availability}
The data that support the findings of this article are not publicly available. The data are available from the authors upon reasonable request.

\appendix
\section{Effect of initial conditions}
\label{app:ic}

\begin{figure}
    \centering
    \includegraphics[width=0.45\textwidth]{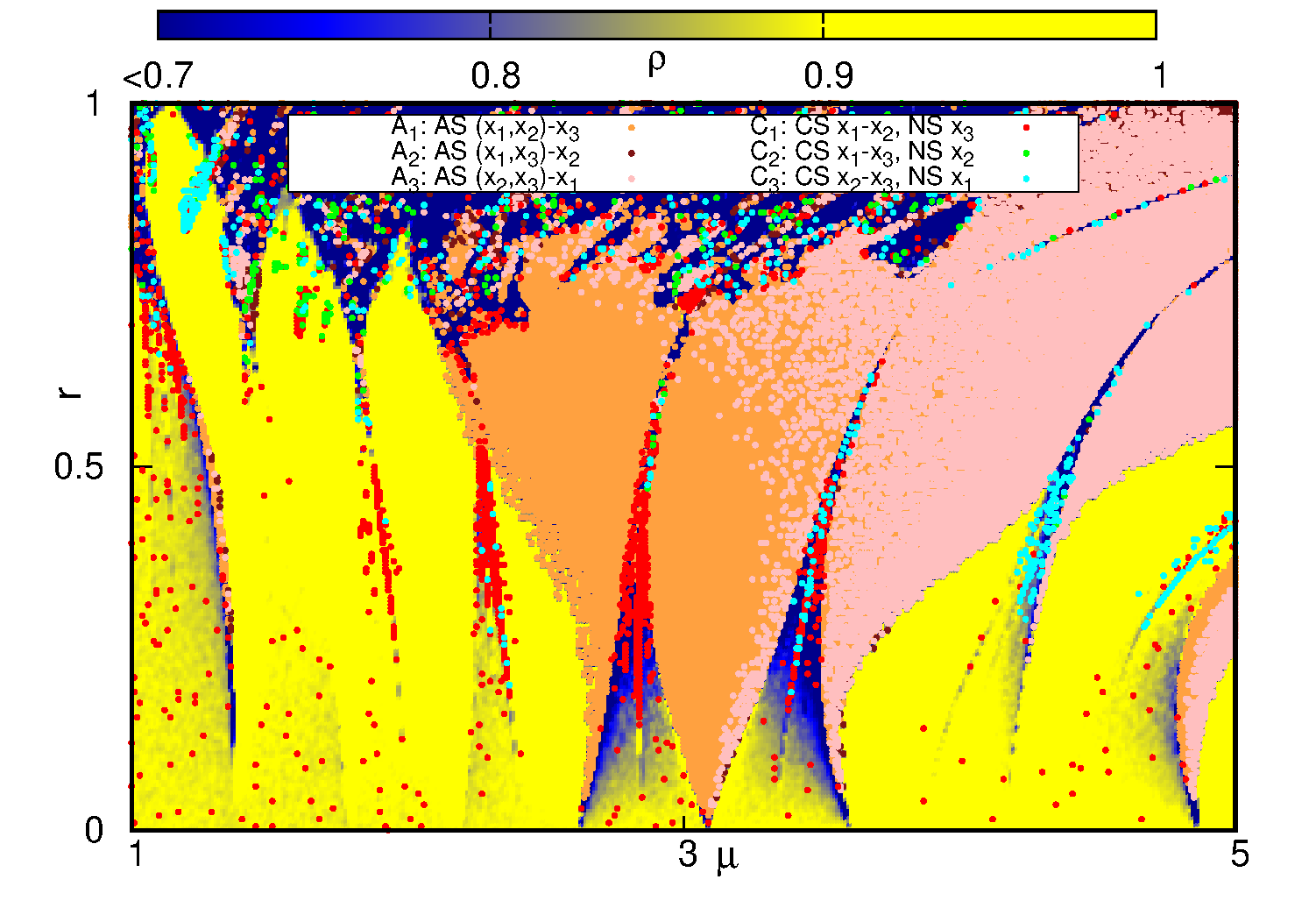}
    \caption{$\mu-r$ plot considering randomized initial conditions for $\delta=0.5$. All other parameters and the color schemes are same as Fig.~\ref{vdp_two_par}.}
    \label{fig:vdp_mu_eta_ic}
\end{figure}

We examine the role of the initial conditions by repeating the $(\mu,r)$ parameter scan with randomly chosen initial states. The resulting synchronization landscape for $\delta=0.5$ is shown in Fig.~\ref{fig:vdp_mu_eta_ic}. A comparison with the fixed-initial-condition diagram (Fig.~\ref{vdp_two_par}(d)) shows that the overall parameter-dependent structure of $\rho$ remains visible. However, the individual collective states become interspersed when the initial conditions are varied. In particular, the three symmetry-related chimera configurations, $\mathrm{C}_1$, $\mathrm{C}_2$, and $\mathrm{C}_3$, occur at numerous parameter points throughout the scan. A similar dependence on the initial state is found for the symmetry-related anti-synchronized configurations also.

This scattered distribution is a consequence of the coexistence of different attracting responses. For the same values of $\mu$ and $r$, changing the initial conditions can place the trajectories in different basins of attraction and hence lead to different collective states. The common drive therefore modifies the nonlinear dynamics in a parameter-dependent manner, while the basin containing the initial state determines which of the coexisting responses is approached.

To quantify this basin dependence, we selected several representative parameter pairs and performed $1000$ independent simulations for each pair using randomly chosen initial conditions. The probability of reaching a collective state $\alpha$ was estimated as
\begin{equation}
p_{\alpha} = \frac{N_{\alpha}}{N_{\mathrm{R}}},
\end{equation}
where $N_{\alpha}$ is the number of realizations classified as state $\alpha$ and $N_{\mathrm{R}}=1000$ is the total number of realizations. The estimated probabilities of the selected collective states are listed in
Table~\ref{tab:ic_prob}.

The finite probabilities of the chimera configurations show that these states are not obtained only from an isolated or specially adjusted initial condition. For example, at $(\mu,r)=(3.5,0.32)$, the probabilities of reaching $\mathrm{C}_1$, $\mathrm{C}_2$, and $\mathrm{C}_3$ are $0.096$, $0.096$, and $0.277$, respectively. Their combined probability is therefore $0.396$.

Similarly, at $(\mu,r)=(3.5,0.35)$, the combined probability of the three chimera configurations is $0.489$. Thus, the chimera states are reached from a finite fraction of randomly sampled initial conditions.

It is also noteworthy that the probabilities of the three chimera configurations are of comparable magnitude. This agrees with the permutation symmetry of the identical oscillators. No oscillator is intrinsically selected as the incoherent unit by the common forcing. The particular realization $\mathrm{C}_1$, $\mathrm{C}_2$, or $\mathrm{C}_3$ is instead selected by the basin of attraction associated with the initial state. 

Similar situations have been observed for time-delayed chaotic system and R\"ossler systems also and has not been included for brevity.

\begin{table}
\caption{Probabilities of states depending upon random initial conditions for $1000$ independent realizations.}
\label{tab:ic_prob}
\centering
\begin{tabular}{l|ccccc}
\hline\hline
Parameters &
$p_{\mathrm{C}_3}$ &
$p_{\mathrm{C}_2}$ &
$p_{\mathrm{C}_1}$ &
$p_{\mathrm{GS}}$ &
$p_{\mathrm{AS}}$\\
\hline
$\mu=3.5$, $r=0.32$~ & $0.277$ & $0.096$ & $0.023$ & $0.014$ & $0.224$\\
$\mu=3.75$, $r=0.4$~ & $0.0$ & $0.0$ & $0.0$ & $0.255$ & $0.0$\\
$\mu=2.0$, $r=0.45$~ & $0.0$ & $0.0$ & $0.0$ & $1.0$ & $0.0$\\
$\mu=3.5$, $r=0.35$~ & $0.281$ & $0.142$ & $0.066$ & $0.022$ & $0.447$\\
$\mu=4.2$, $r=0.36$~ & $0.003$ & $0.05$ & $0.04$ & $0.139$ & $0.0$\\
\hline\hline
\end{tabular}
\end{table}

\section{R\"{o}ssler system}
\label{app:ros}

\begin{figure}
\centering
\includegraphics[width=0.48\textwidth]{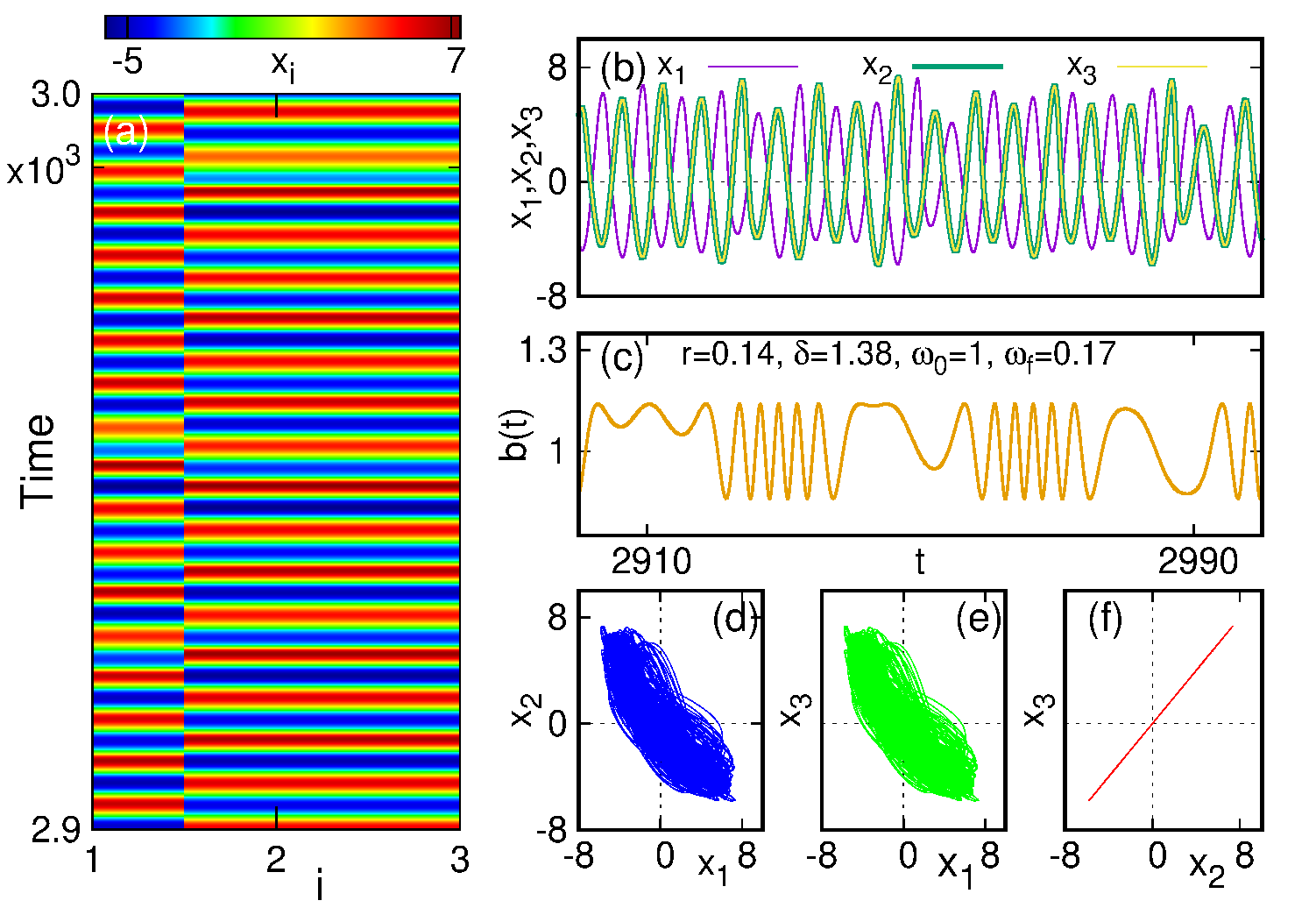}
\caption{Minimal chimera state in the frequency-modulated R\"{o}ssler system. (a) Spatiotemporal evolution of the three oscillators. (b) Time series of $x_i$. (c) Frequency-modulated forcing signal. (d),(e) Phase portraits in the $x_1$–$x_2$ and $x_1$–$x_3$ planes showing desynchronization. (f) Phase portrait in the $x_2$–$x_3$ plane showing synchronization. Parameters: $a = 0.2$, $b = 0.2$, $c = 3.2$, $r = 0.14$, $\delta = 1.38$, $\omega_0 = 1.0$, and $\omega_f = 0.17$.}
\label{ros_fm_sptmp}
\end{figure}

To further examine the generality of the proposed mechanism, we consider a network of three R\"{o}ssler system subjected to frequency-modulated forcing. The governing equations are given by
\begin{subequations}\label{roseq}
\begin{align}
\dot{x}_i &= -y_i - z_i, \label{roseqa}\\
\dot{y}_i &= x_i + a y_i, \label{roseqb}\\
\dot{z}_i &= b + z_i\big(x_i - c(1 + r \sin(\vartheta_0))\big), \label{roseqc}
\end{align}
\end{subequations}
where $a \in \mathbb{R}^+$, $b \in \mathbb{R}^+$, and $c \in \mathbb{R}^+$ are system parameters. The phase $\vartheta_0$ follows Eq.~\eqref{phieq}. Hence, the frequency modulated parameter becomes $c(t)=c\left[1 + r\sin\left\{\omega_0\left(t - \frac{\delta}{\omega_f}\cos(\omega_f t)\right)\right\}\right]$. Here we use, $a=b=0.2$, $c=3.2$, and the modulation parameters: $r = 0.14$, $\delta = 1.38$, $\omega_0 = 1.0$, and $\omega_f = 0.17$. As initial conditions we use: $(x_1(0), y_1(0), z_1(0))=(0.9,0.4,0.1)$, $(x_2(0), y_2(0), z_2(0))=(0.6,0.3,0.4)$, and $(x_3(0), y_3(0), z_3(0))=(0.3,0.2, 0.05)$ throughout the investigation.

The R\"{o}ssler system exhibits dynamical behavior consistent with that observed in the previous models. In particular, the coexistence of synchronized and desynchronized oscillators under common frequency-modulated forcing is clearly evident, confirming the emergence of minimal chimera states. For brevity, we present here a representative case illustrating the spatiotemporal evolution, time series, forcing signal, and phase-space projections (Fig.~\ref{ros_fm_sptmp}), which collectively demonstrate the same qualitative features discussed earlier.

\section{Data Acquisition using Data Acquisition System}
\label{app:dac}

For recording and visualizing the signals obtained from the electronic circuit, we employ a National Instruments data acquisition system (DAQ NI USB-6351, 8 inputs, 2 outputs, maximum sampling rate 1.25 MS/s) interfaced with a desktop computer through the LabVIEW environment \cite{labview2014}.

\begin{figure}
    \centering
    \includegraphics[width=0.48\textwidth]{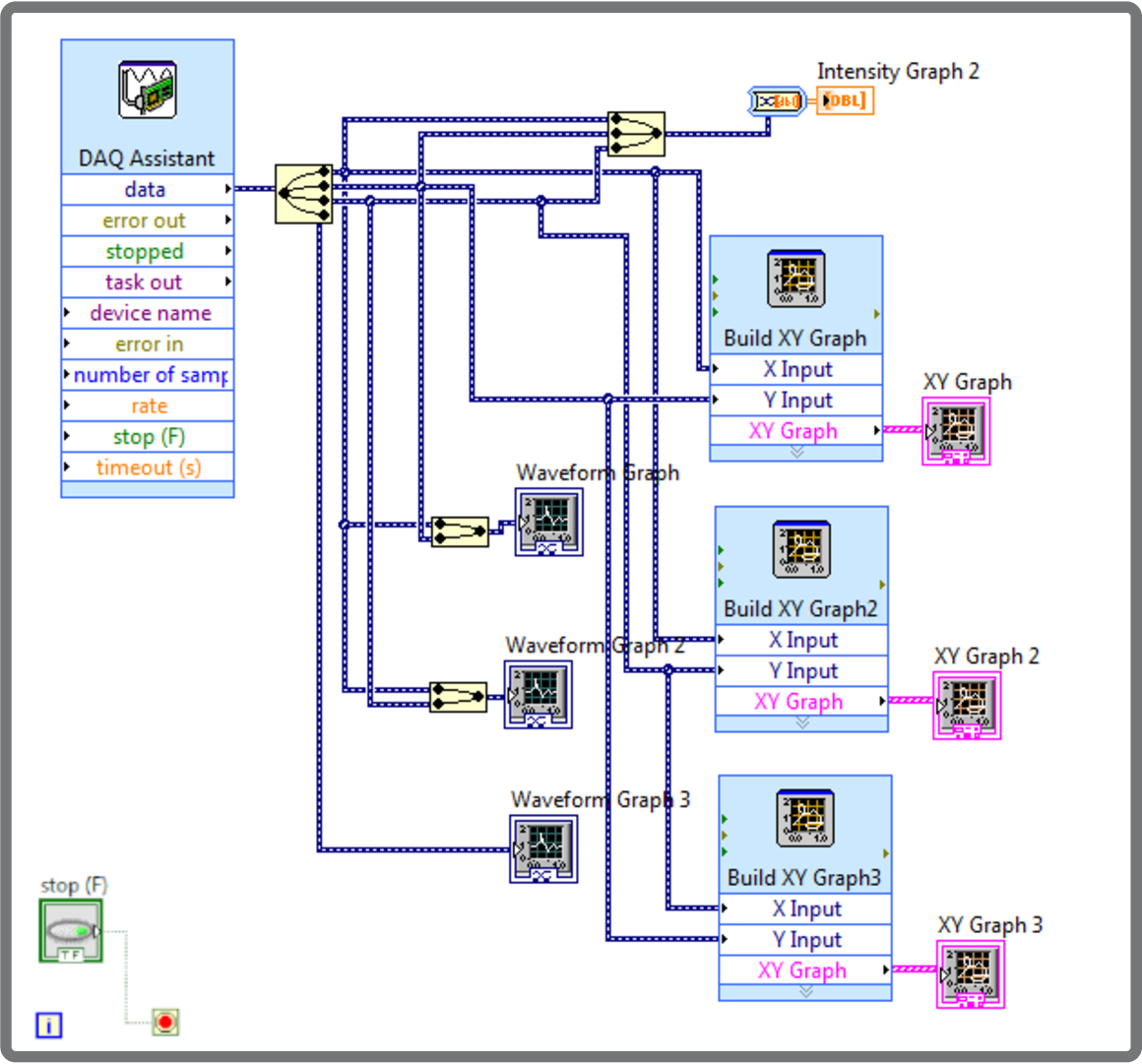}
    \caption{Block diagram of the LabVIEW-based data acquisition and visualization system used in the experiment.}
    \label{block_diag}
\end{figure}

The LabVIEW program consists of two main components: a block diagram, which defines the data acquisition and processing workflow, and a front panel, which provides real-time visualization of the measured signals. The required modules for data acquisition are incorporated using the DAQ Assistant, while graphical outputs such as time series and spatiotemporal plots are generated through appropriate visualization blocks.

Figure~\ref{block_diag} shows the block diagram of the LabVIEW program used in the experiment. This configuration enables synchronized acquisition of multiple voltage signals and their simultaneous visualization, which is essential for identifying synchronization and chimera states in the system.

\section{Frequencies of the oscillators}
\label{app:freq}

We calculate the long-time effective frequencies of the three oscillators. For the limit-cycle system, the effective angular frequency of the $i$-th oscillator has been evaluated from the accumulated phase as \cite{pikovsky2001synchronization}
\begin{equation}\label{apomeq}
\Omega_i^{\mathrm{eff}} = \lim_{T_{\mathrm{obs}}\rightarrow\infty} \frac{\theta_i(t_0+T_{\mathrm{obs}})-\theta_i(t_0)}{T_{\mathrm{obs}}},
\end{equation}
after discarding the initial transient. 
For the chaotic time-delayed and R\"ossler systems, a uniformly rotating phase is not naturally available. We therefore estimate the mean angular frequency from the number of completed oscillation cycles over a sufficiently long observation interval, following the usual mean-phase-velocity characterization of oscillatory and chimera dynamics
\cite{bera2016chimera}. The effective angular frequency is evaluated as
\begin{equation}\label{apomeqeff}
\Omega_i^{\mathrm{eff}} = \frac{2\pi N_i}{t_{i,N_i}-t_{i,0}},
\end{equation}
where $N_i$ denotes the number of completed oscillation cycles during
the observation interval.

The calculated frequencies are tabulated in Table~\ref{tab:effective_frequency}. For comparison, the corresponding autonomous oscillator without
frequency-modulated driving has also been calculated.

\begin{table*}[t]
\caption{Effective angular frequencies for the representative dynamical states under autonomous and frequency-modulated conditions.}
\label{tab:effective_frequency}
\centering
\begin{tabular}{lcccc}
\hline\hline
System &
$\Omega_{\mathrm{aut}}$ &
$\Omega_1^{\mathrm{eff}}$ &
$\Omega_2^{\mathrm{eff}}$ &
$\Omega_3^{\mathrm{eff}}$ \\
\hline
Van der Pol & $0.60754$~~ & $0.62805$~~ & $0.62804$~~ & $0.62808$ \\
Time-delayed chaotic system~~~~~ & $0.82323$~~ & $1.06655$~~ & $1.06298$~~ & $1.06298$ \\
R\"ossler system & $1.08973$~~ & $1.09507$~~ & $1.09492$~~ & $1.09492$ \\
\hline\hline
\end{tabular}
\end{table*}

These results clarify that the frequency-modulated signal does not simply impose its carrier frequency on the oscillators. Instead, the external modulation changes their long-time mean oscillation rates relative to the corresponding autonomous systems. The effect depends on the underlying dynamics. For the van der Pol system, the three effective frequencies are nearly identical, although the oscillators do not all exhibit complete phase locking. The time-delayed chaotic system shows a small but finite frequency splitting between the incoherent oscillator and the synchronized pair. For the Rössler system, the mean frequencies are again very close, despite the dynamically distinct behavior of the incoherent oscillator. These results indicate that a minimal chimera cannot, in general, be identified solely from a difference in long-time mean frequencies. Equal or nearly equal effective frequencies do not necessarily imply bounded relative phases or complete synchronization.


\bibliography{minimal_chim_fm}

\end{document}